\documentclass[preprint]{aastex}

\usepackage{natbib}
\usepackage{array}
\usepackage{color}
\usepackage[dvipsnames]{xcolor}

\newcommand{\Rsun}{${\mathrm R}_{\odot}$}

\newcommand{\argmin}{\arg\!\min}
\newcommand{\review}[1]{#1}

\begin{document}

\title{3D electron density distributions in the solar corona during solar minima: assessment for more realistic solar wind modeling}
\shorttitle{3D electron density distributions in the solar corona during solar minima}
\shortauthors{J. de Patoul et al.}


\author{Judith de Patoul\altaffilmark{1,2}, 
Claire Foullon\altaffilmark{1}, 
and Pete Riley\altaffilmark{3}}

\affil{\altaffilmark{1} Centre for Geophysical and Astrophysical Fluid Dynamics -- University of Exeter, Exeter, Devon EX4, UK}
\email{j.depatoul@exeter.ac.uk, c.foullon@exeter.ac.uk}
\affil{\altaffilmark{2} Laboratoire d'Astrophysique de Marseille, UMR 7326, CNRS and Aix-Marseille Universit\'e, 38 Rue Fr\'ed\'eric Joliot-Curie, F-13388 Marseille cedex 13, France}
\affil{\altaffilmark{3} Predictive Science, 9990 Mesa Rim Road, Suite 170, San Diego, CA 92121, USA}
\email{rileype@saic.com}


\begin{abstract}
Knowledge of the electron density distribution in the solar corona put constraints on the magnetic field configurations for coronal modeling and on initial conditions for solar wind modeling.
We work with polarized SOHO/LASCO-C2 images from the last two recent minima of solar activity (1996--1997 and 2008--2010), devoid of coronal mass ejections. 
\review{The goals are to} derive the 4D electron density distributions in the corona by applying a newly developed time-dependent tomographic reconstruction method \review{and to compare the results between the two solar minima and with two magnetohydrodynamic models}. 
First, \review{we confirm that the values of the density distribution in thermodynamic models are more realistic than in polytropic ones.} The tomography provides more accurate distributions in the polar regions, 
and we find that the \review{density in tomographic and thermodynamic solutions} varies with the solar cycle in both polar and equatorial regions. 
Second, we find that the highest-density structures do not always correspond to the predicted large-scale heliospheric current sheet or its helmet streamer but can follow the locations of pseudo-streamers.  
\review{We deduce that} tomography offers reliable density distributions in the corona, reproducing the slow time evolution of coronal structures, without prior knowledge of the coronal magnetic field over a full rotation.  
Finally, we suggest that the highest-density structures show a differential rotation well above the surface depending on how they are magnetically connected to the surface. Such valuable information on the rotation of large-scale structures could help to connect the sources of the solar wind to their in situ counterparts in future missions such as \textit{Solar Orbiter} and \textit{Solar Probe Plus}.

\end{abstract}



\section{Introduction}
\label{sec_intro}

The distribution of the magnetic field generated in the solar interior and connected into the solar wind influences most coronal phenomena, including large-scale and slowly evolving coronal structures. 
The coronal density distribution can serve as a tracer of the configuration of the magnetic field (shape and general morphology rather than field strength), since the coronal plasma is frozen into the field \citep[for a review see, e.g.,][]{Wiegelmann2014A&ARv}. 
Of particular interest are the observations of streamers and pseudo-streamers, referring to structures associated with large magnetic loops that separate coronal holes of opposite polarity,
and twin loop arcades that separate coronal holes of the same polarity respectively \citep{Wang2007ApJ_pseudostreamers}. 
Another example is the study of magnetic structures above the solar polar regions, where the measurements of the line-of-sight (LOS) magnetograms are generally less reliable owing to the larger viewing angle with the magnetic field. The observed electron density distributions in coronal holes and polar plumes \citep{Barbey2008SoPh, dePatoul2013SoPh} could provide a better understanding of how the flux emergence near the equator affect the magnetic field configuration at the pole  \citep{dePatoul2013AA}.
Finally, an accurate determination of the ambient coronal electron density provides a better estimation of the mass and the propagation of coronal mass ejections (CMEs) \citep{Vourlidas2000, Feng2015ApJa, Feng2015ApJb}.
In particular, the density is important for calculating the compression ratio of CME-driven shocks and the Alfv\'en Mach number, which has important implications for the localization of particle acceleration sites and hence space weather forecasts \citep{Bemporad2011, Chen2014}.

The first proposed empirical approach to obtain the electron density from remote sensing observations
was an inversion method using measurements from eclipses in polarized white light, with the assumption that the coronal electron density is axisymmetric \citep{vandeHulst1950}. \cite{Saito1977} used this method to calculate electron densities from polarized brightness (pB) observations obtained by \textit{Skylab} coronagraph data  during the declining phase of the solar cycle from 1973 May to 1974 February. A good agreement of the density values was found using SOHO/LASCO-C2 data during 1998 February \citep{Hayes2001ApJ} and 1996 December \citep{QuemeraisAnA2002}.
Empirical methods to obtain the full 3D density distribution are given by solar rotational tomography (SRT). SRT has been specifically developed for optically thin structures and uses LOS-integrated coronal images from multiple viewpoints taking advantage of solar rotation.
White-light images of the K-corona, where the radiation is dominated by Thomson scattering, can be used to reconstruct density from 1.5~\Rsun\ up to 6.5~\Rsun\ using images from the LASCO-C2 or \textit{STEREO}/COR1 \citep[e.g.,][]{Frazin2000, Barbey2013SoPh, Kramar2014SoPh, Peillon2014}. When the sources for a tomographic inversion are EUV images, both density and temperature can be reconstructed by applying differential emission measure tomography  \citep{Frazin2009ApJ}. However, even in the best cases, only reconstruction close to the surface from about 1.03~\Rsun\ to 1.25~\Rsun\ can be obtained.
An alternative physics-based approach to obtain a quantitative 3D density distribution is given by magnetohydrodynamic (MHD) models, which provide the global configuration of the magnetic field and the plasma parameters (i.e., density, temperature, and velocity) in the corona \citep{RileyJGR2001, RileyApJ2006, Lionello2009}.

Here we determine the electron density distribution in the corona during the two previous solar minima: 1996--1997 (solar cycle number 22/23) and 2008--2010 (solar cycle number 23/24).
In section~\ref{sec_meth}, we determine the 4D electron density distribution ($N_{e}$) from a newly developed  time-dependent tomographic method. We look at the general morphology of the density structures in the empirical model from tomography and compare with a simple potential field source surface (PFSS) model and more advanced MHD models. 
In section~\ref{sec_resu}, we contrast the density values found by tomography and the ones predicted by MHD models; especially, we discuss 
(1) the temporal and radial profiles of the density, 
(2) the location of the helmet streamer and pseudo-streamer, and
(3) the presence of a differential rotation of the structures in the corona.

\section{Determination of the electron density distribution}
\label{sec_meth}
\subsection{$N_{e}$ from Tomography}
\label{sec_meth_Tomo}

\begin{figure}
\includegraphics[width=0.4\textwidth, trim=3.5cm 1.8cm 4.7cm 1.3cm, clip]{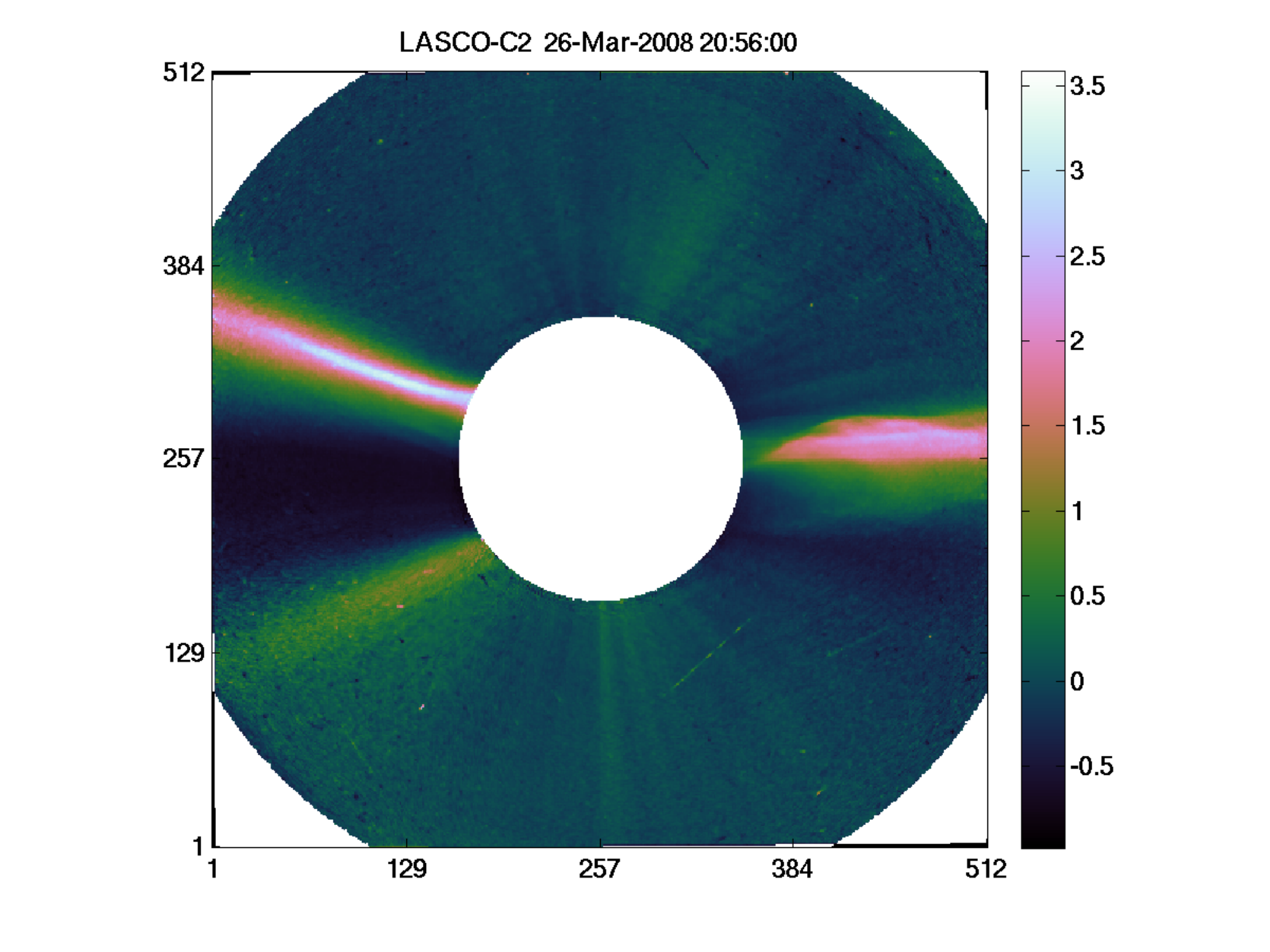}\\[1ex]
\includegraphics[width=0.4\textwidth, trim=3.5cm 1.8cm 4.7cm 1.3cm, clip]{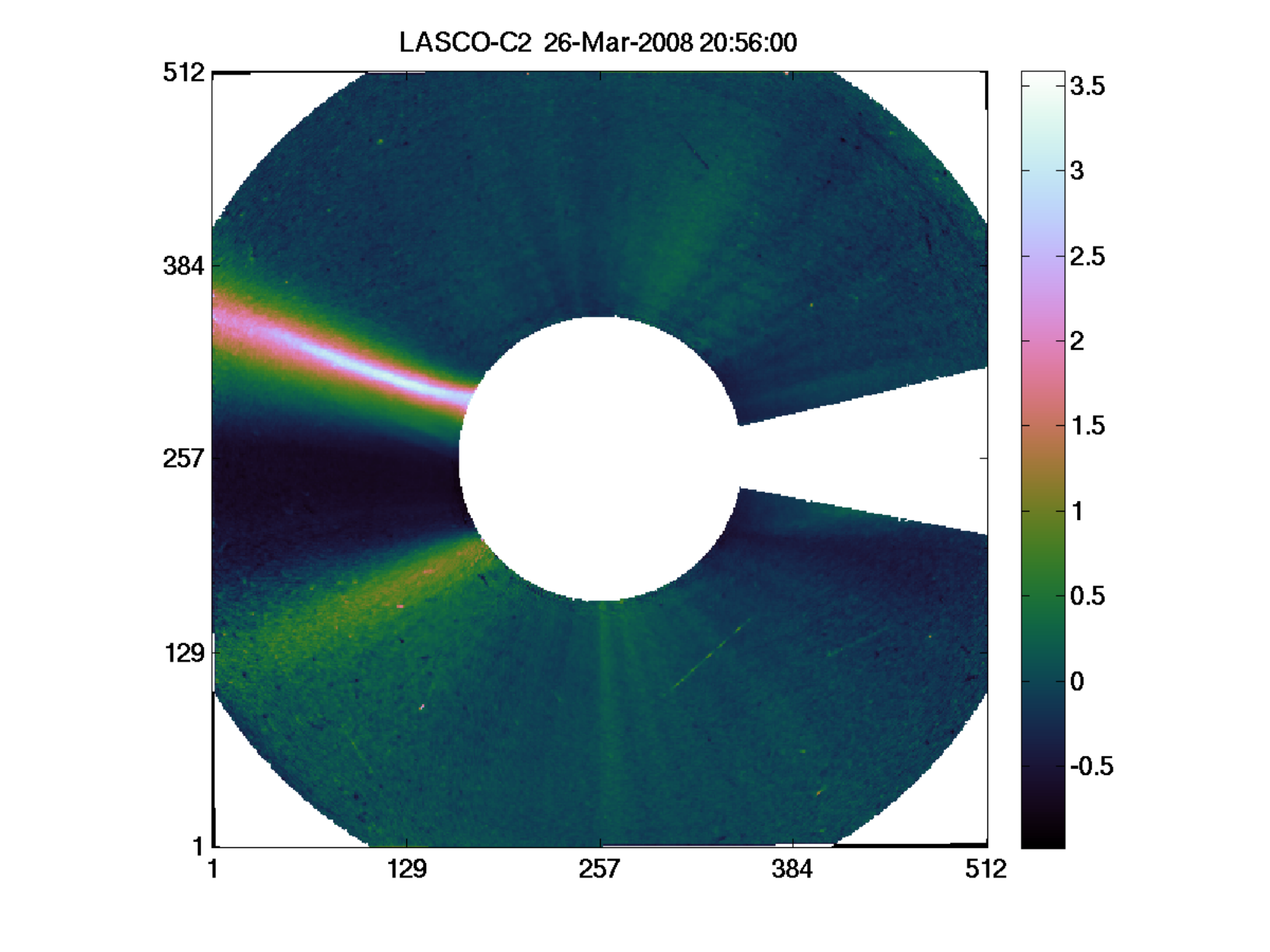}\\
\caption{SOHO/LASCO-C2 pB image taken on the 2008 March 26  at 20:56UT. The field of view is shown from 2.5~\Rsun\ to 6.5~\Rsun.
A contrast enhancement has been applied to increase the intensity in the radial direction.
Top: pB image with a CME observed. 
Bottom: same image with the angular portion of the CME, which is excluded from the tomographic procedure.\label{fig_lasco}}
\end{figure}

Since 1996 the SOHO/LASCO-C2 coronagraph has continuously produced sets of white-light and polarized images of the solar corona with a field of view ranging from about 1.5~\Rsun\ up to 6.5\Rsun\ \citep{BruecknerSoPh1995}.
To determine the electron density distribution ($N_e$) in the corona,
we use the pB images that are extracted from the total brightness LASCO-C2 images pre-processed as described by \cite{llebaria_2006, gardes2013,LamyJGR2014}. 
The resulting pB images are dominated by the electron-scattered K corona, which is known to be strongly polarized \citep{Billings1966}, and not contaminated by the dust-scattered F corona, 
which is essentially unpolarized at low heights and has been removed during the calibration.
The intensity measured in pB images, $I_{\rm pB}$, observed from a view direction at a rotation angle, $\vartheta$, of the Sun relative to the observer's longitude, is the integration of the electron density, $N_e$, along the LOS direction, $\vec{e}_{\rm LOS}(\vartheta)$,
\begin{equation}
I_{\rm pB} (\rho,\vartheta)
= \int_{\rm LOS}
N_{e} \bigg(\vec{r} \big(l\ \vec{e}_{\rm LOS}(\vartheta)\big) \bigg) \
K \bigg(  \vec{r} \big(l\ \vec{e}_{\rm LOS}(\vartheta)\big), \ \rho\ \vec{e}^{\perp}_{\rm LOS}(\vartheta) \bigg)  \ {\rm d}l,
\label{eq_IpB}
\end{equation}
where $\vec{e}^{\perp}_{\rm LOS}$ is the vector unit orthogonal to $\vec{e}_{\rm LOS}$,
$\rho$ is the distance from the Sun center to $\vec{e}_{\rm LOS}$ and
$\vec{r}$ is the radial vector.
The Thomson scattering function, $K$, is defined for a point source of luminosity, $4\pi L$, by  \citep{Frazin2010}:
\begin{equation}
K = \frac{3\sigma_{e}}{16}\frac{L}{\rho^2} \sin^4\Theta,
\label{eq_ThomScatt}
\end{equation}
where $\Theta$ is the scattering angle defined by $\sin\Theta = \frac{\rho}{\|\vec{r}(l\ \vec{e}_{\rm LOS}(\vartheta))\|}$ and
$\sigma_e$ is the Thomson scattering cross section for a single electron.
A typical example of a pB image is shown in Figure~\ref{fig_lasco} (top panel), where a background subtraction has been applied to enhance the intensity along the radial direction.
In this work, we consider coronal heights above 2.5~\Rsun\ to avoid artifacts due to diffraction surrounding the occulter. 
The pB image shown in Figure~\ref{fig_lasco} (top panel) was taken when a CME occurred at a position of 271$^\circ$. This CME had an angular width of 114.1$^\circ$ and traversed the corona from 2.5~\Rsun\ to 7.5~\Rsun\ in approximately 2.5 hr \citep{Boursier2006}.
Solar rotational tomographic methods cannot resolve fast temporal changes, and important artifacts are produced in the reconstructions.
To minimize this effect, we remove the CMEs from the pB images.
\cite{Morgan2010CME} proposed a method for separating CMEs from the quiescent corona in white-light coronagraph images based on the fact that the large-scale structures are close to radial, whilst CMEs are highly nonradial.
Here we consider CMEs listed in the CACTus \citep{RobbrechtAnA2004} and the ARTEMIS \citep{Boursier2006} catalogs that have an intensity larger than 0.8$\times 10^3$W sr$^{-1}$ m$^{-2}$. 
Using the position angle and angular width from these catalogs, we simply exclude the angular portion of the pB image affected by the CME from the tomographic reconstruction procedure (Figure~\ref{fig_lasco}, bottom panel).

The electron density is obtained by inverting equation (\ref{eq_IpB}) using SRT.
We use the newly developed time-dependent tomographic method, which has been elaborated and described by \cite{Peillon2014, Peillon2014Poster}.
The method involves spatio-temporal regularization (Kalman filters) to mitigate the slow temporal variation of the corona
and assumes a nearly solid rotation of the Sun of 27.2753 days corresponding to the Carrington rotation.
It requires a continuous set of view directions uniformly distributed over half a rotation, with a minimum cadence of one pB image per day, i.e., a number of $n_I\geq$13 images for a given tomography reconstruction. 
The corona is divided into a spherical grid $(r, \phi, \theta; t)$ with a size of ($60\times60\times120\times n_I$), covering the heliocentric distances from 2.5 to 8.5~\Rsun. 
To assess the robustness and accuracy of the technique, 
the method has been tested using a set of 14 projected images of a time-dependent MHD volume as ``observations''. The result could successfully reproduce the slow time-varying dynamic of the model.
The estimated density distribution, $\tilde{\bf{x}}$, is constructed on the grid cells by solving the following least-squares minimization problem:
\begin{equation}
\tilde{\bf{x}} = \argmin_{\bf{x}\geqslant 0} \Big\{
\left\| \bf{y}-\bf{A}\bf{x}\right\|^{2}_{2}
+ \lambda_S^{2}   \left\|  \bf{R}_S \bf{x} \right\|^{2}_{2}
+ \lambda_T^{2}   \left\|  \bf{R}_T \bf{x} \right\|^{2}_{2}
+ \lambda_C^{2}   \left\|  \bf{R}_C \bf{x} \right\|^{2}_{2}
\Big\}.
\label{eq_tomo}
\end{equation}
The vector $\bf{y}$ contains the intensity measured in each pixel from the set of pB images over half a rotation, i.e., the $I_{\rm pB} (\rho_{ij},\vartheta)$ defined in Equation (\ref{eq_IpB}) with $\vartheta \in [0,2\pi]$ and $\rho_{ij}$ giving the position of the pixel in the image.
The vector $\bf{x}$ contains $N_e$ values defined in the spherical grid $(r, \phi, \theta; t)$.
$\bf{A}$ is a diagonal-like matrix composed of blocks of projection matrices 
that are determined by the geometry and the physics of the problem, i.e., the relation between the volume element in $\bf{x}$ and the LOS-related pixel in the pB image defined by Thomson scattering function (\ref{eq_ThomScatt}).
The matrices $\bf{R_S}, \bf{R_T}$ and $\bf{R_C}$ in equation (\ref{eq_tomo}) are the spatio-temporal regularization terms, which introduce a prior knowledge of the solution. 
This regularization minimizes the effects of the noise, the limited number of pB images available, and the unavoidable temporal change in the corona.
The spatial regularization matrix, $\bf{R_S}$, described in \cite{Frazin2007}, is a second derivative of the angular spherical coordinates $\theta$ and $\phi$,  multiplied by $r^{-1}$ to reduce the radial distance noise.
The temporal regularization matrix, $\bf{R_T}$, is a first derivative to enforce smoothness between two successive views of the Sun.
The co-rotating regularization matrix, $\bf{R_C}$, is acting jointly in the space-time domain.
Its purpose is to prevent the reconstruction from concentrating material in the vicinity of the
plane of the sky (containing the Sun's center). 
This is a plane that rotates in the Carrington coordinate system, and it is always orthogonal to the observer's LOS.
The regularization parameters, $(\lambda_S,\lambda_T,\lambda_C)$, are estimated by minimizing the normalized root means square error of the time-dependent 3D MHD model and its reconstruction ($\lambda_{S}=2.2\ 10^{-6}$, $\lambda_{T}=1.7\ 10^{-6}$ and $\lambda_{C}=0.2\ 10^{-6}$).
Further details about the method and the construction of these regularization operators can be found in \cite{Peillon2014}; see, \review{in particular discussion on the use of the temporal regularization, including examples of 3D and 4D tomographic reconstruction}.

\begin{figure}
\begin{tabular}{m{0.1cm} m{0.58\textwidth} m{0.30\textwidth}}
(a)  &
\includegraphics[width=0.58\textwidth, trim=0 1.36cm 0 0, clip]{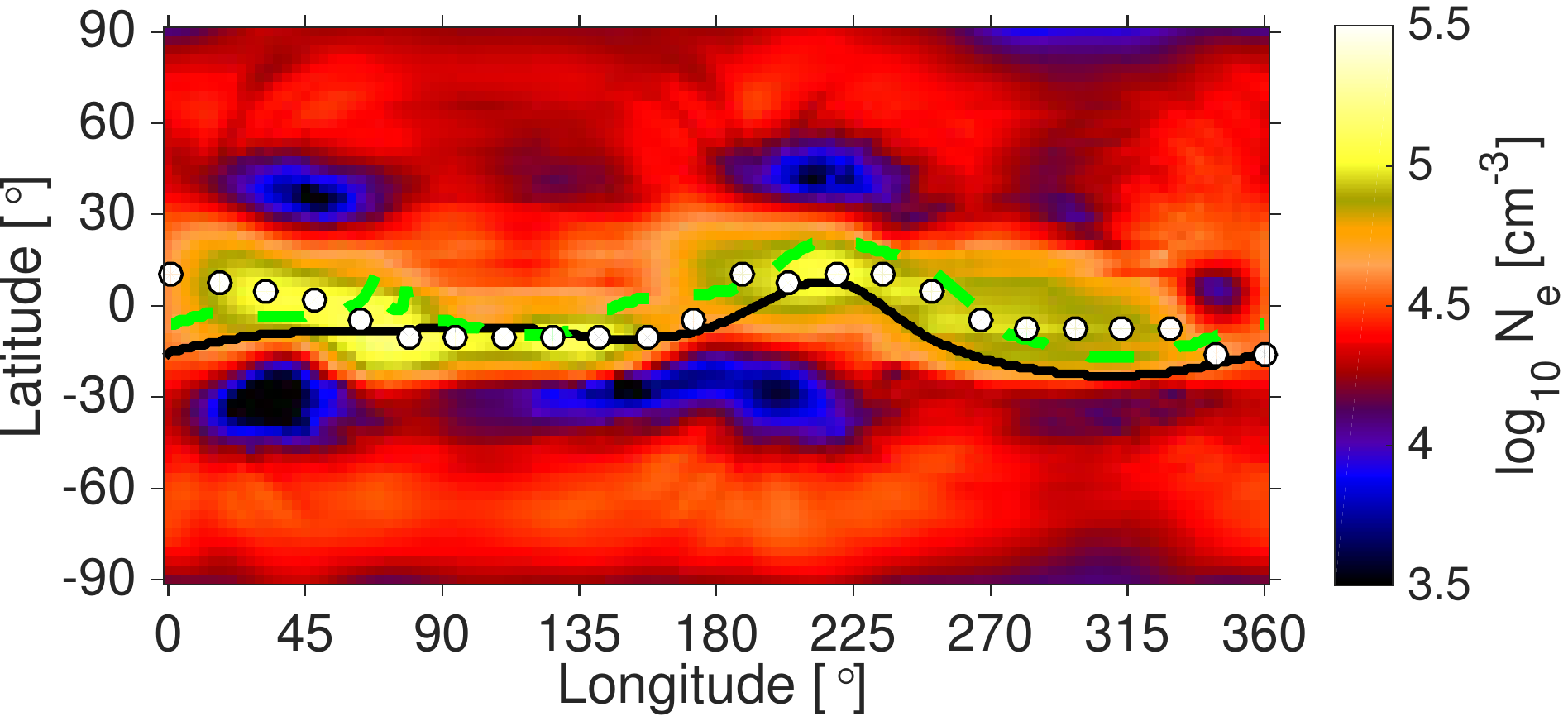} &
\includegraphics[width=0.30\textwidth, trim=0 1.55cm 0 0, clip]{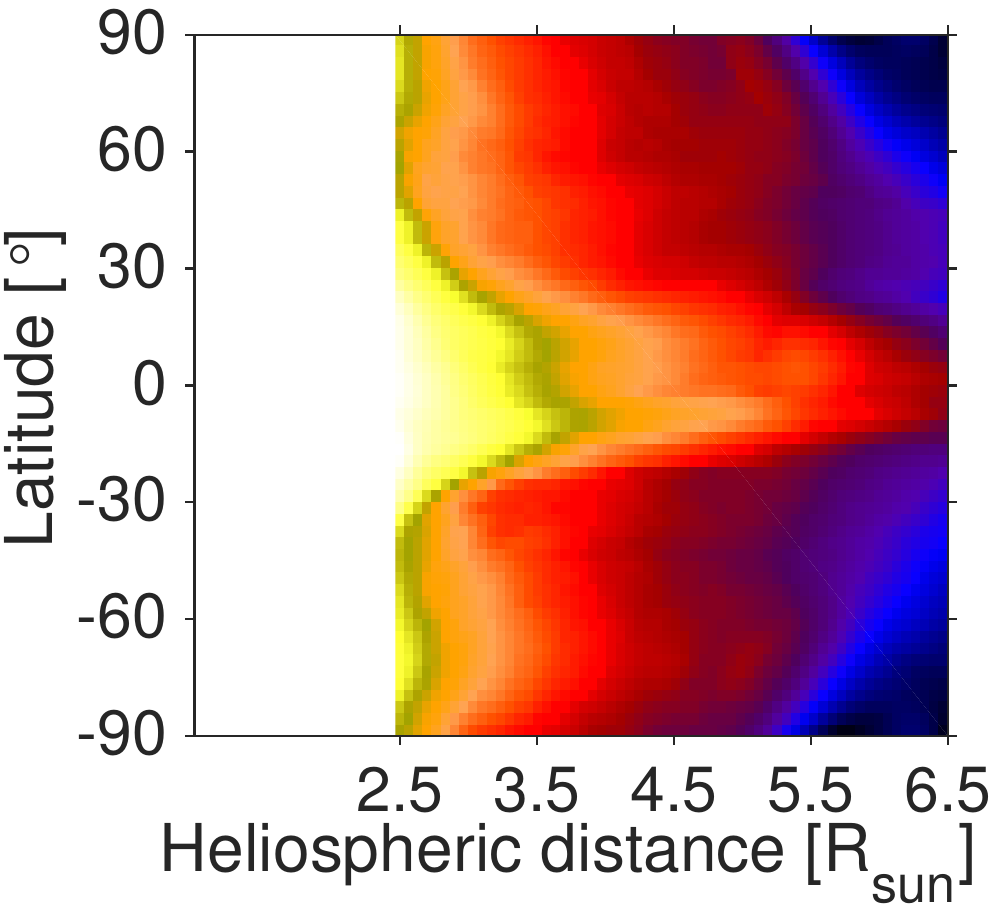} \\
(b)  &
\includegraphics[width=0.59\textwidth, trim=0 1.36cm 0 0, clip]{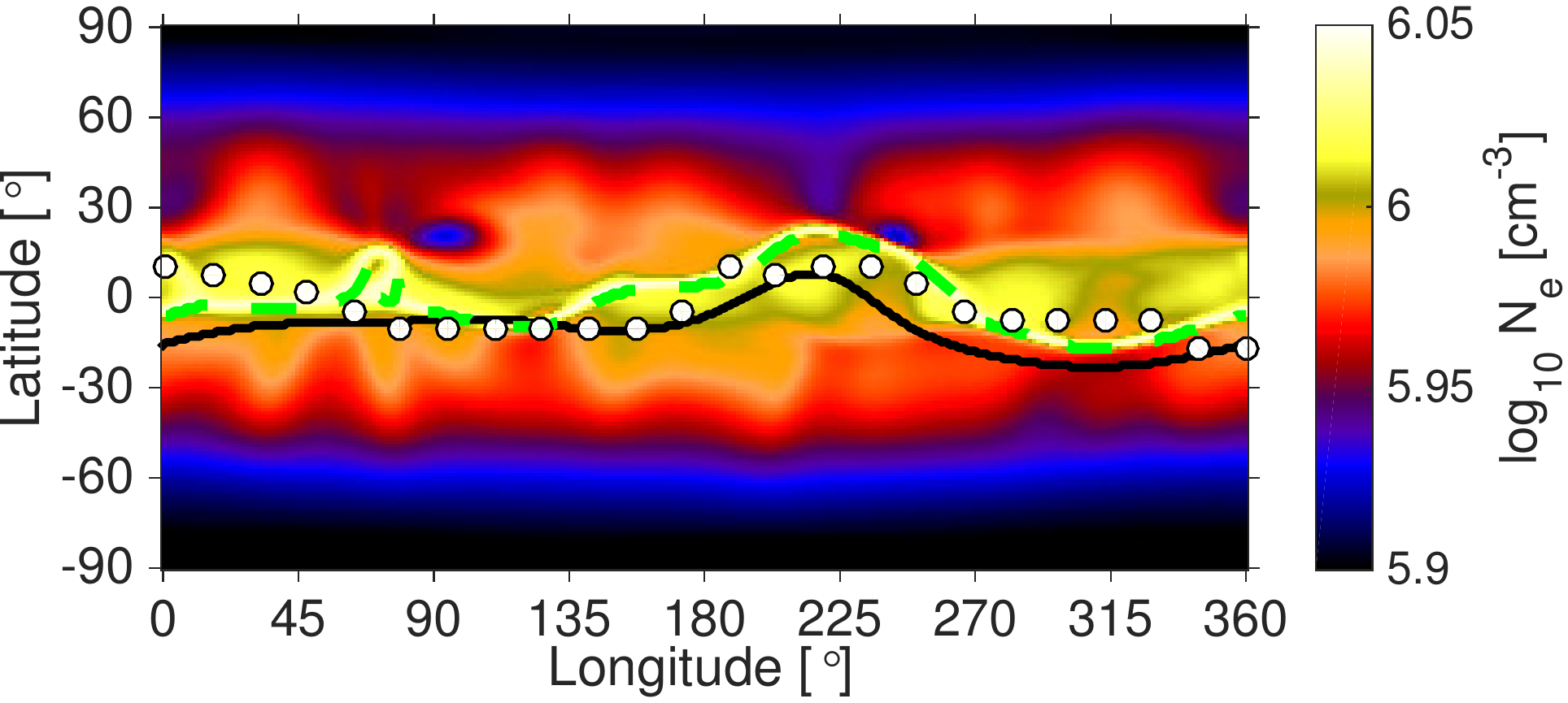} &
\includegraphics[width=0.30\textwidth, trim=0 1.55cm 0 0, clip]{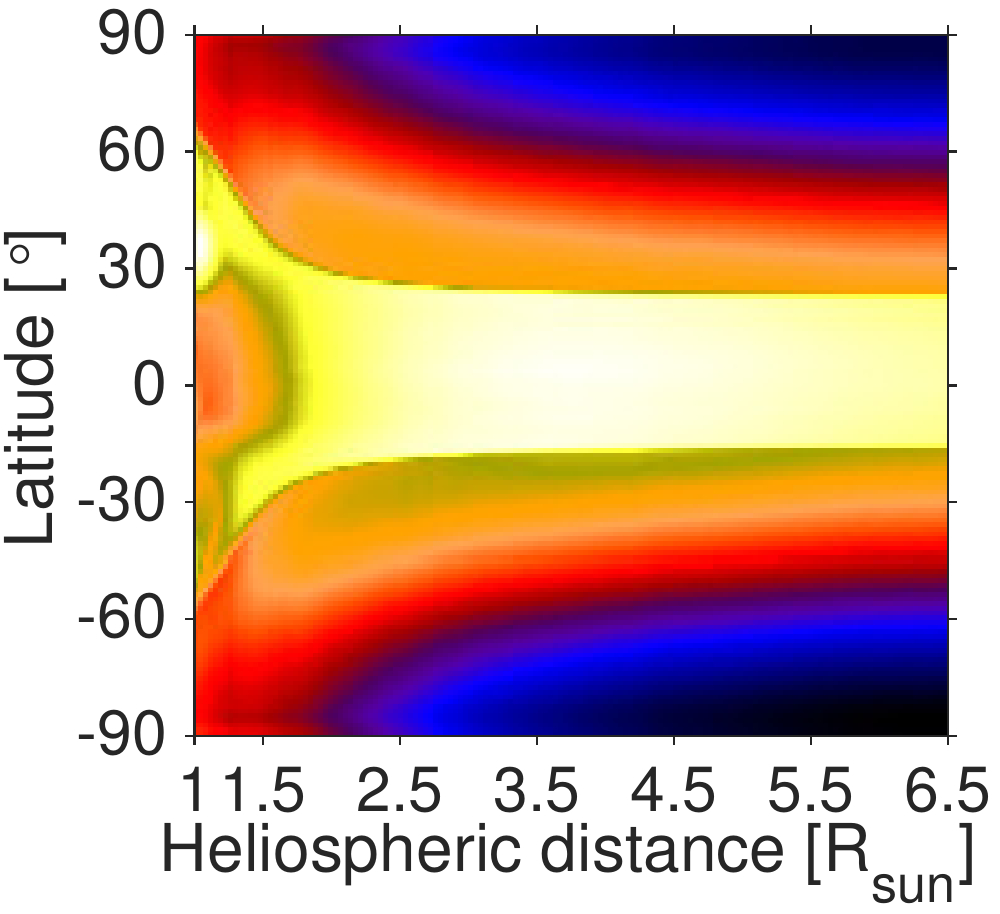} \\
(c)  &
\includegraphics[width=0.58\textwidth, trim=0 1.36cm 0 0, clip]{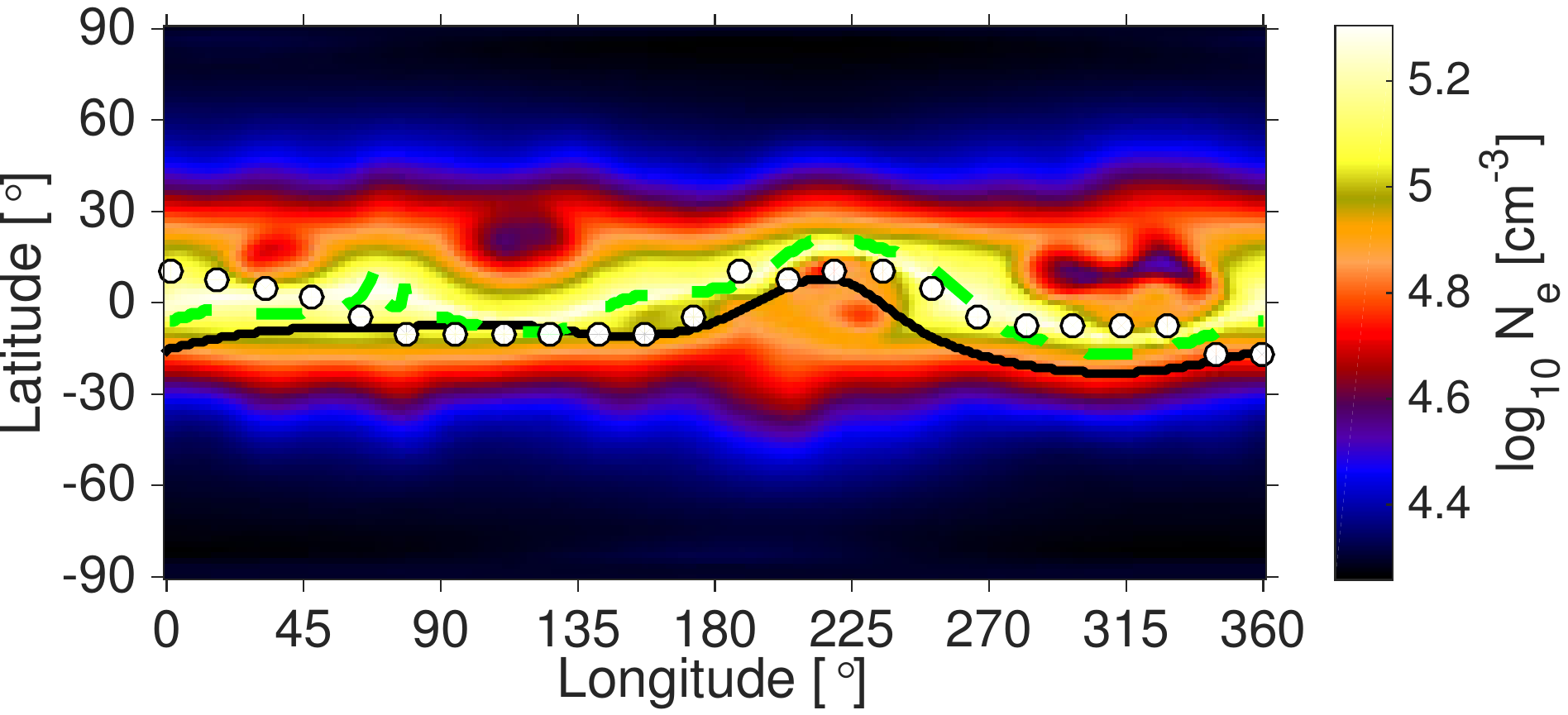} &
\includegraphics[width=0.30\textwidth, trim=0 1.55cm 0 0, clip]{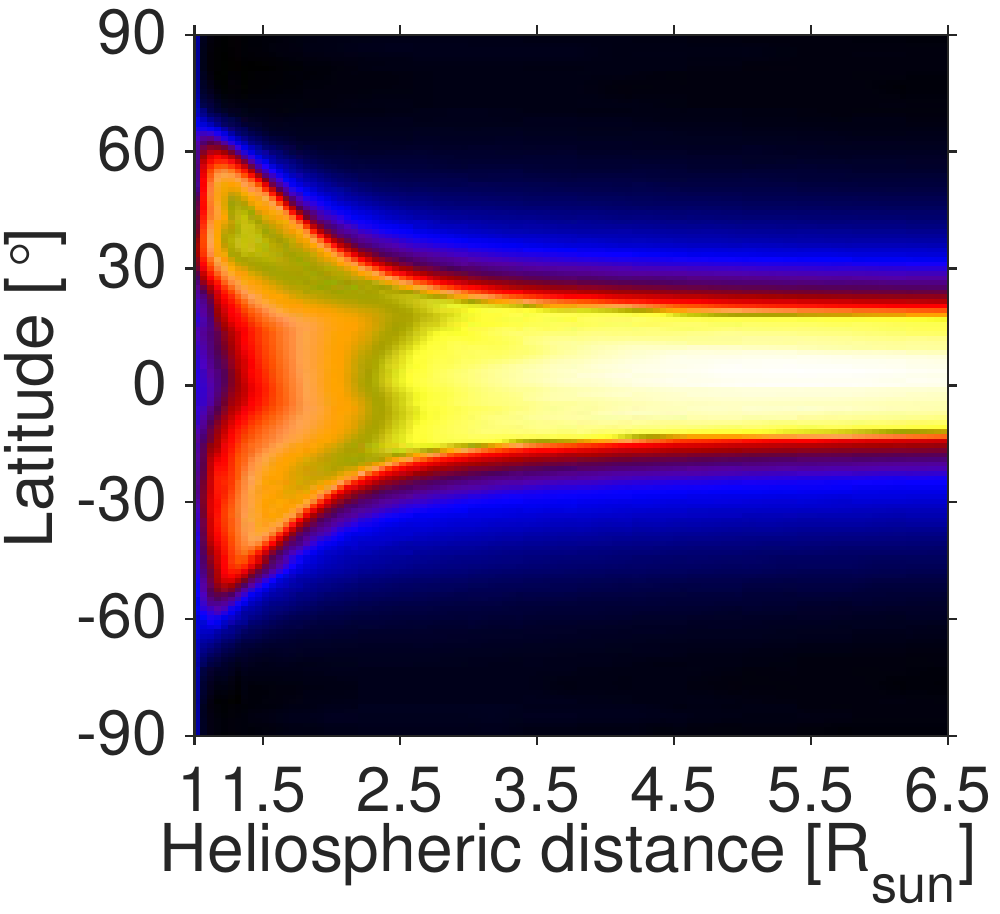} \\
(d)  &
\includegraphics[width=0.575\textwidth, trim=0 0 0 0, clip]{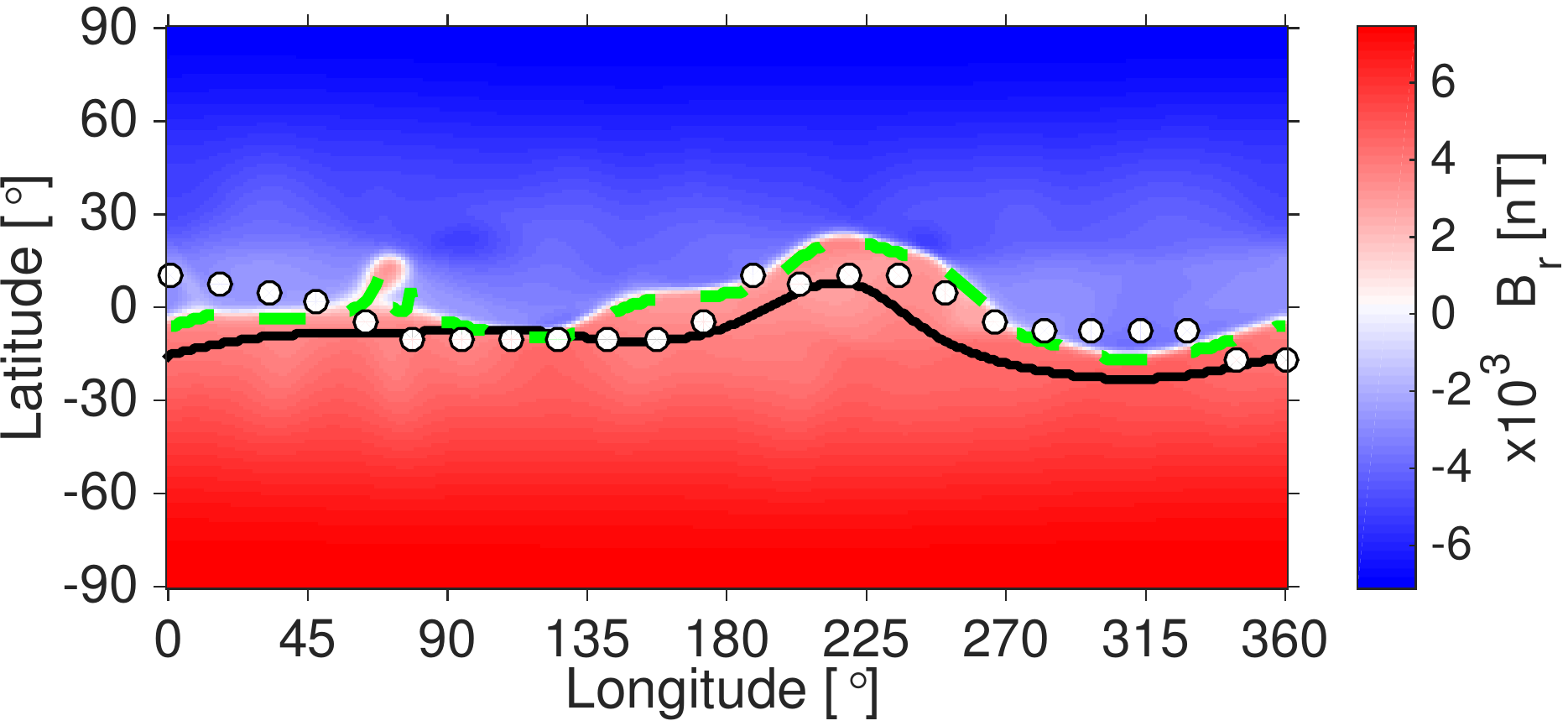} &
\includegraphics[width=0.30\textwidth, trim=0 0 0 0, clip]{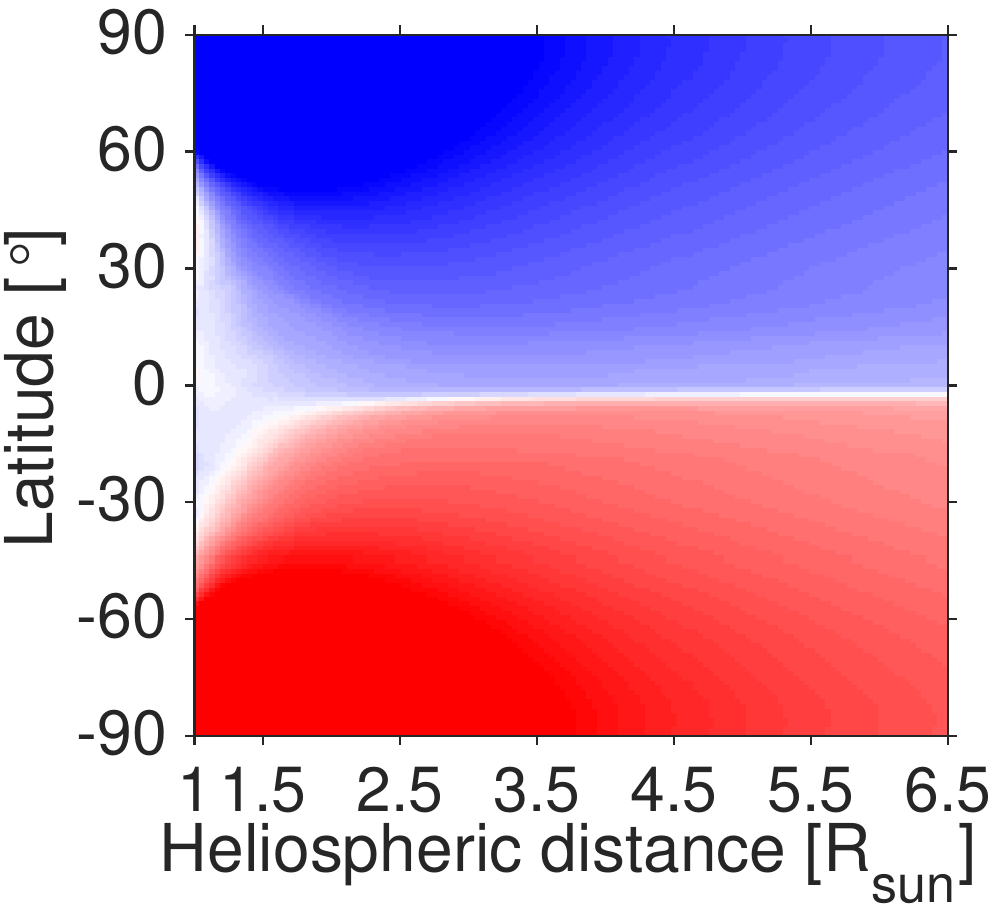}
\end{tabular}
\caption{(a) $N_e$ from tomography (2008 November 21 to December 4). 
\review{MHD solutions for Carrington rotation 2077: (b) polytropic pMHD/$N_e$, (c) thermodynamic tMHD/$N_e$ and (d) polytropic pMHD/$B_r$.}
The left side shows the longitude--latitude map at 3.5~\Rsun. 
See text for an explanation of the plotted lines. 
The right panel shows the latitude--radial maps obtained by integrating over the longitudes in tomographic and MHD/$N_e$ solutions, and by a median over the longitudes in the pMHD/$B_r$ solution (radial contrast enhancement has been applied).  \label{fig_tomo_predsci_2077}}
\end{figure}

A full 4D reconstruction is performed every 4 days, provided that a minimum of 13 pB images are available.
During 1996--1997, several data gaps are present for which the tomography was not carried out.
Panel (a) of Figure~\ref{fig_tomo_predsci_2077} shows a typical result from tomography during a relatively quiet period of the solar activity when the number of CMEs is reduced.
It was obtained using 14 pB images from 2008 November 21 to December 4, which is included in the Carrington rotation 2077.
The left panel of Figure~\ref{fig_tomo_predsci_2077}~(a) shows the 2D longitude--latitude map at 3.5~\Rsun\ centered on 2008 December 2. The right panel shows the latitude--radial average map constructed by integrating over the longitudes (a radial contrast enhancement has been applied).
It helps to represent the extent to which the helmet streamer spreads over the latitudes during this particular period.
Panel (a) of Figure~\ref{fig_tomo_predsci_2097} shows another result from tomography in the later phase of the extended solar minimum, when solar activity has started to increase.
It was obtained using 15 pB images from 2010 June 6 to 20, during the Carrington rotation 2097.
The latitudinal positions of the maximum of density evaluated for each longitude in the tomographic reconstruction are indicated by the white dots. 
Some voxels near the higher-density structure have a density value close to zero,
for example, Figure~\ref{fig_tomo_predsci_2077}~(a), the region at longitude [30$^{\circ}$, 34$^{\circ}$] and latitude [-29$^{\circ}$, -32$^{\circ}$]. 
These zero-density artifacts are usually caused by the unavoidable rapid change in the corona. 
Indeed, the inverse problem can set a negative value to account for an unexplained variation of intensity in the data from a single viewpoint \citep{Barbey2013SoPh}. This could also be caused by remaining instrumental artifacts in a pB image.

\subsection{$N_{e}$ from MHD models}
\label{sec_meth_MHDmod}

\begin{figure}
\begin{tabular}{m{0.1cm} m{0.58\textwidth} m{0.30\textwidth}}
(a)  &
\includegraphics[width=0.58\textwidth, trim=0 1.36cm 0 0, clip]{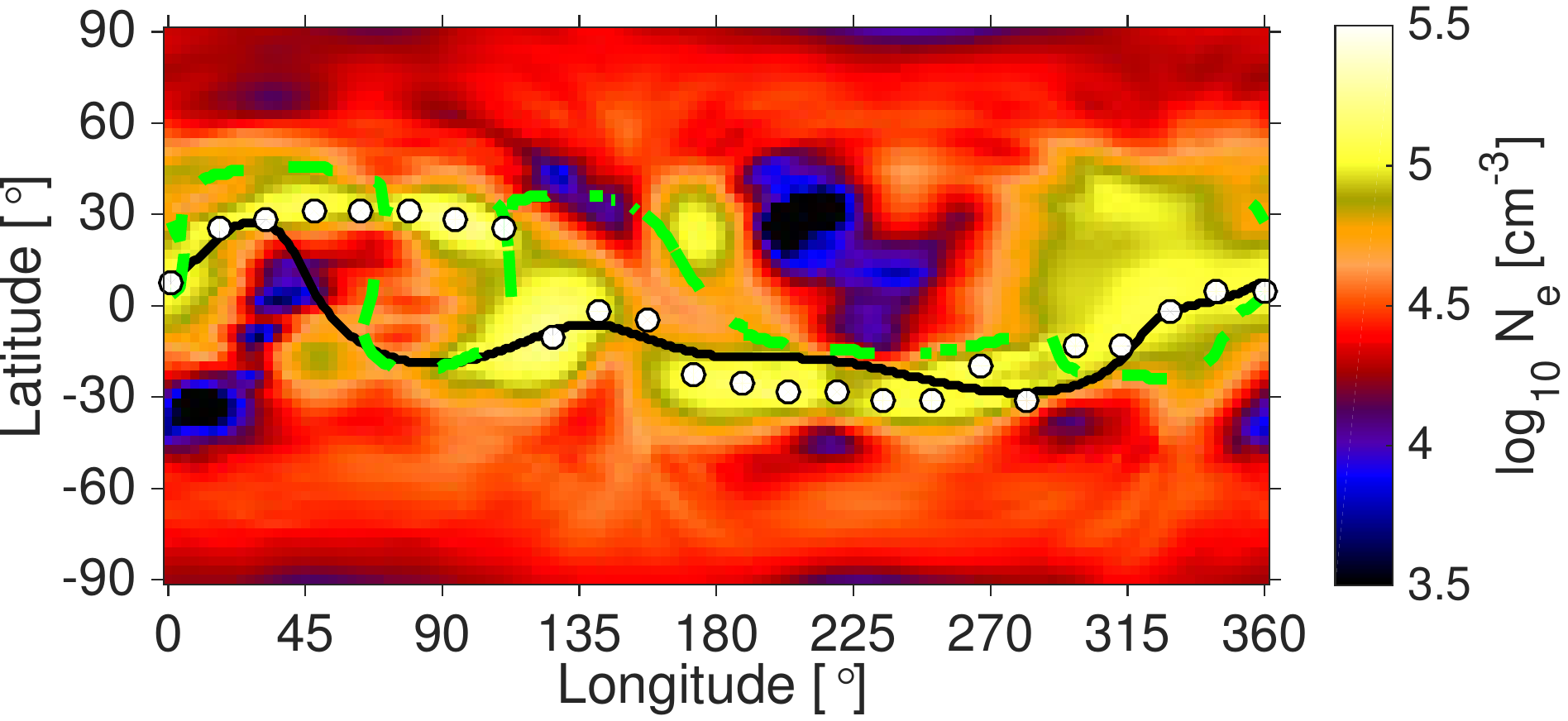} &
\includegraphics[width=0.30\textwidth, trim=0 1.55cm 0 0, clip]{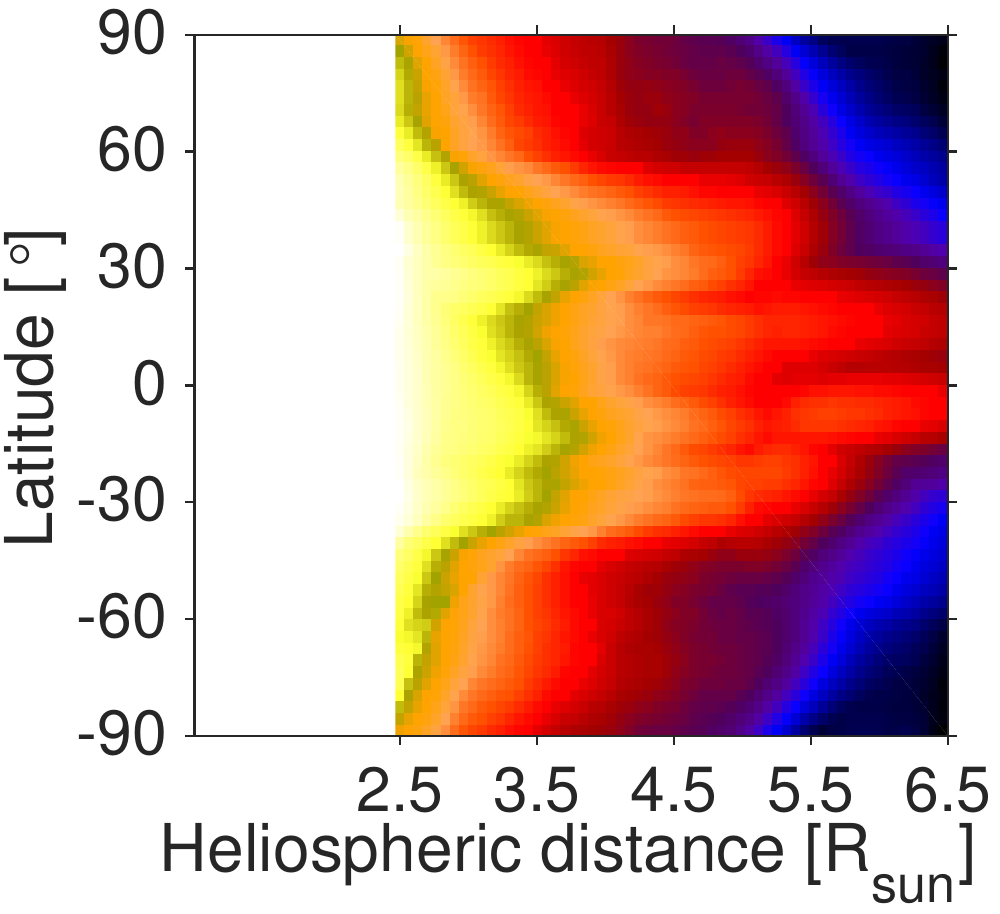} \\
(b)  &
\includegraphics[width=0.59\textwidth, trim=0 1.36cm 0 0, clip]{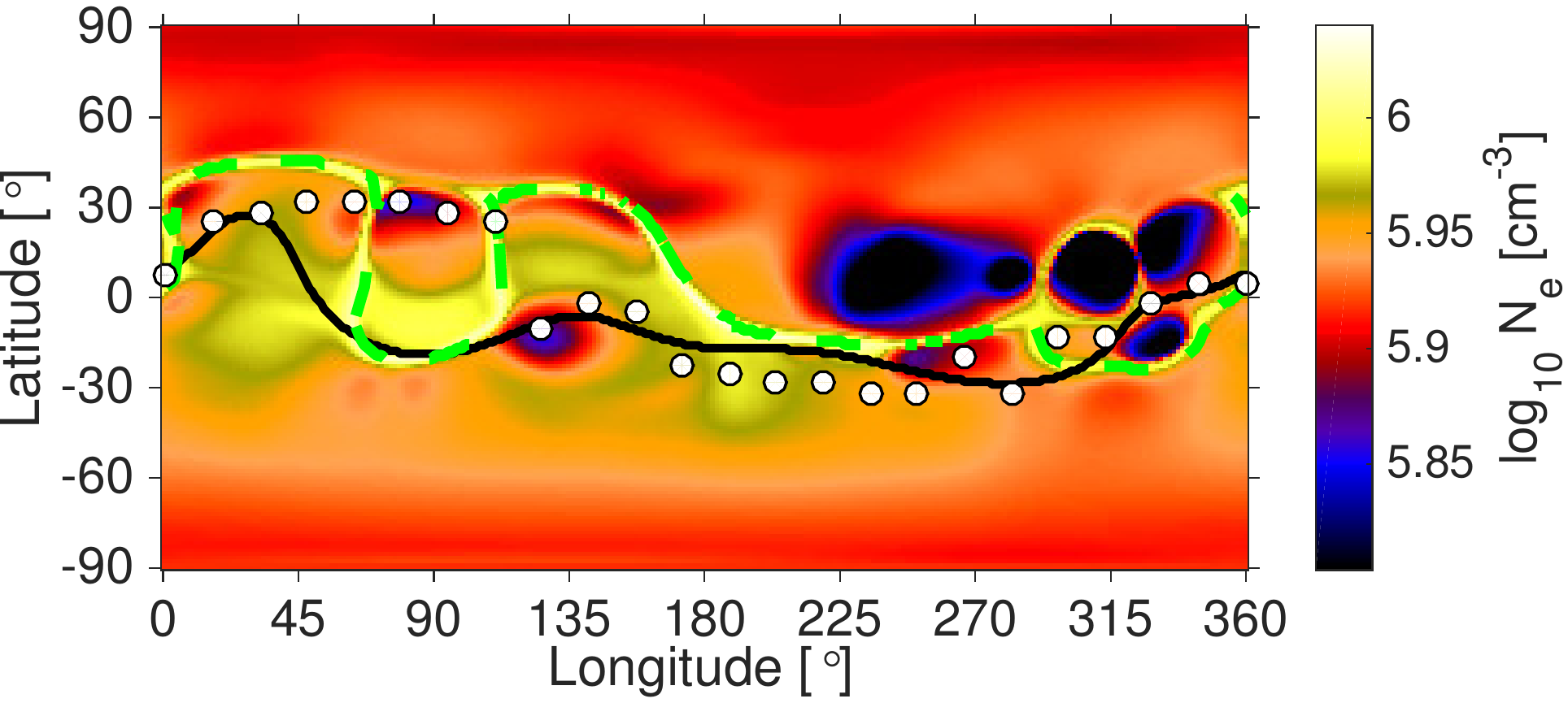} &
\includegraphics[width=0.30\textwidth, trim=0 1.55cm 0 0, clip]{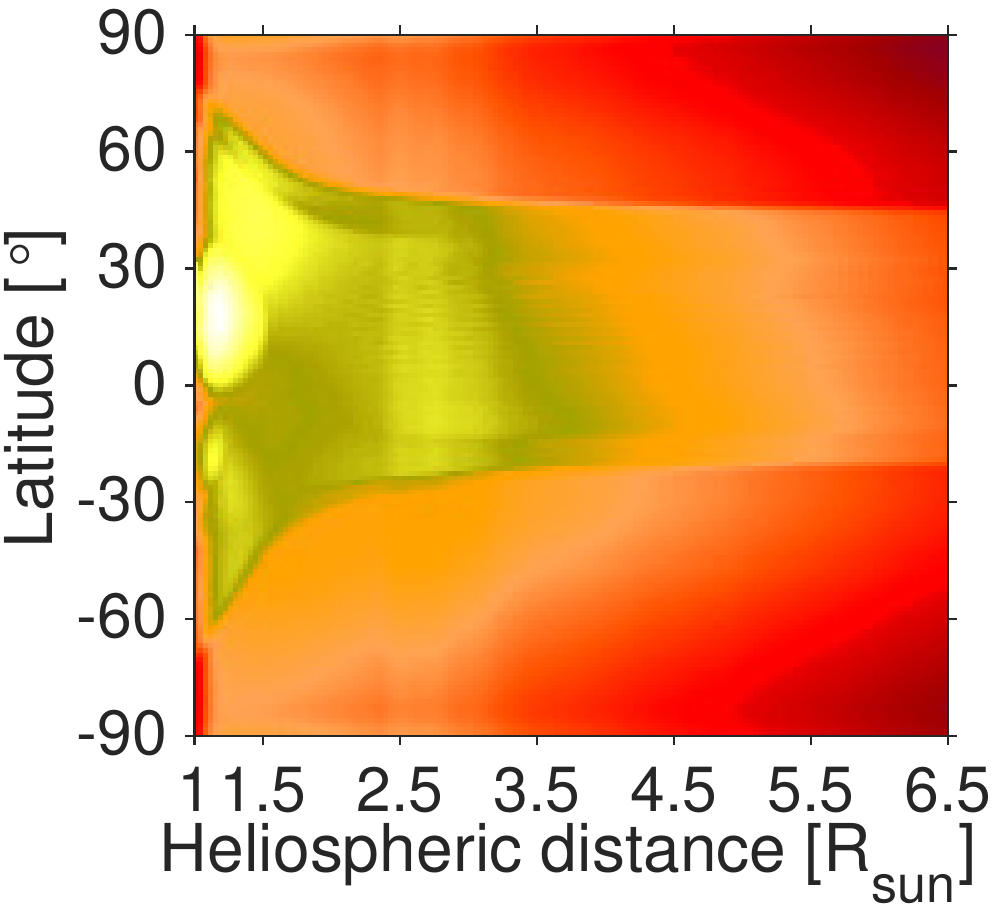} \\
(c)  &
\includegraphics[width=0.575\textwidth, trim=0 0 0 0, clip]{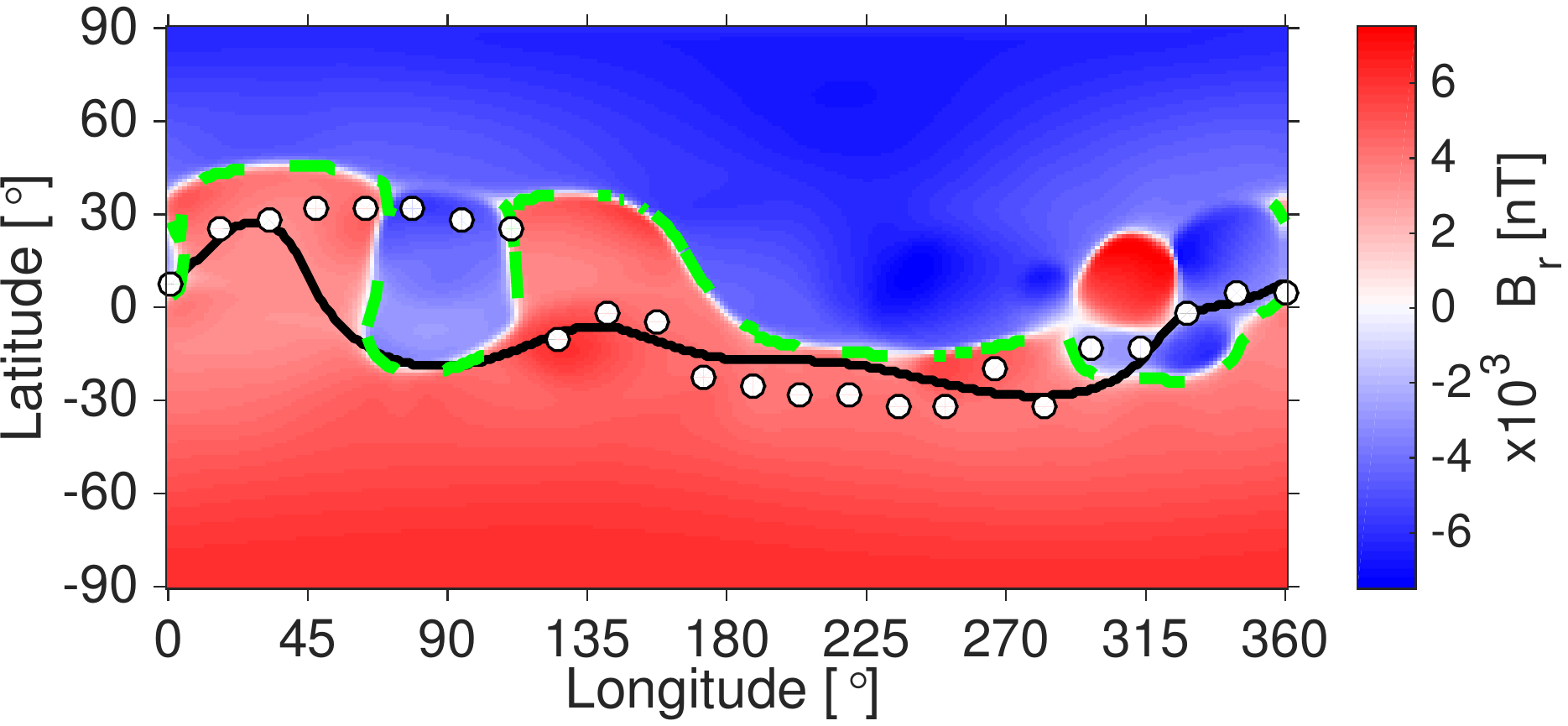} &
\includegraphics[width=0.30\textwidth, trim=0 0 0 0, clip]{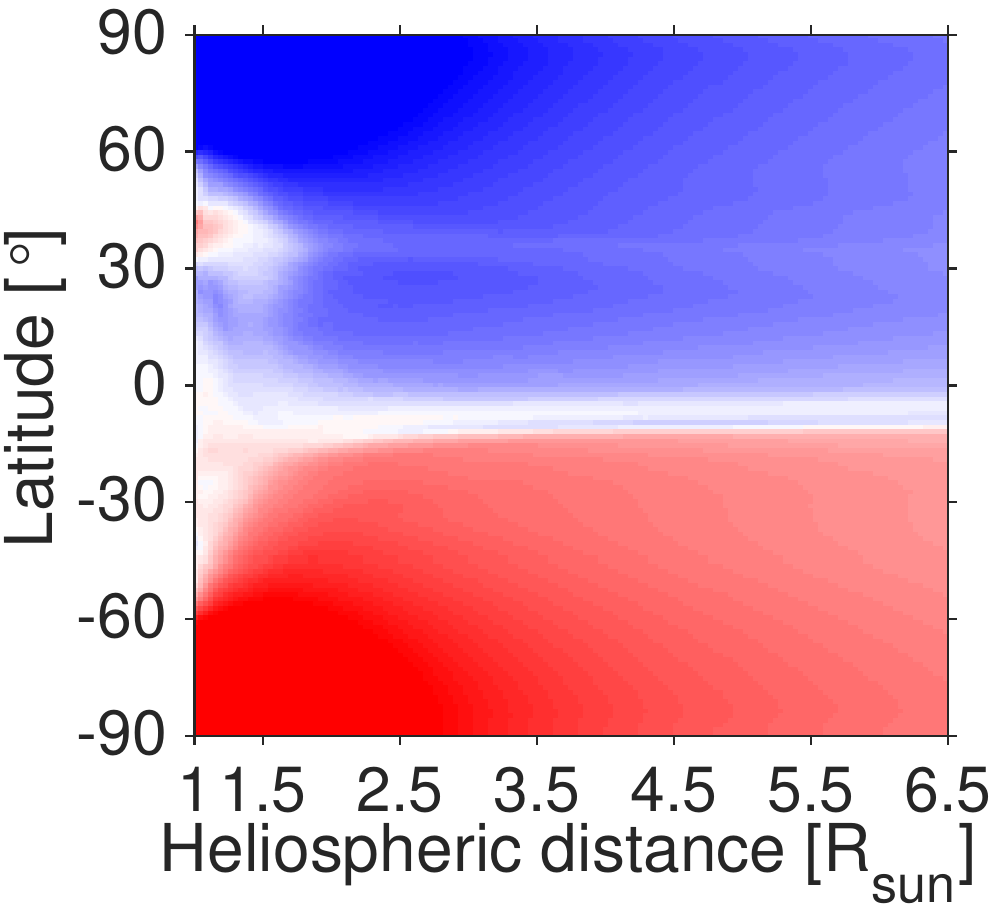}
\end{tabular}
\caption{Same as Figure~\ref{fig_tomo_predsci_2077}. (a) Tomographic solution from 2010 June 6 to 20. Polytropic MHD solution for Carrington rotation 2097 in (b) and (c). \label{fig_tomo_predsci_2097}}
\end{figure}

The PFSS model is a simple and popular current-free model capable of reproducing the basic coronal magnetic field configuration. It requires only the synoptic maps of LOS photospheric magnetic field component as lower boundary, and it assumes that all field lines become radial at the upper boundary (the source surface) at about 2.5--3.5~\Rsun.
The global magnetic field configuration predicted by the PFSS model can be used as a proxy of the density distribution in the corona. In particular, the neutral line at the source surface, which separates the large-scale opposite-polarity regimes of the coronal magnetic field, is often used to locate the heliospheric current sheet (HCS) and the helmet streamer. The PFSS/HCS calculated for a source surface at 2.5~\Rsun\ is displayed as the black line in Figures~\ref{fig_tomo_predsci_2077} and \ref{fig_tomo_predsci_2097}.

A more complex and elaborate way to predict the magnetic field configuration and the density distribution in the corona is to employ global MHD models. We use solutions from MHD models developed by the group at Predictive Science \citep[][see online, \url{www.predsci.com}]{RileyJGR2001, RileyApJ2006, Lionello2009}. 
For the lower boundary condition, the models use the radial component of the magnetic field provided by the observed LOS measurements of \textit{SOHO}/MDI magnetograms and uniform characteristic values for the plasma density and temperature. 
It assumes also that the electron and proton density are equal.
In the polytropic MHD model, the energy equation is approximated by a simple adiabatic energy equation with a polytropic index $\gamma=1.05$. 
Since this approximation significantly simplifies the problem and reduces the time necessary to complete a simulation, its solutions can be obtained more routinely and are available between 1~\Rsun\ and 30~\Rsun\ for all the Carrington rotations under study.
This model reproduces well the geometrical and topological properties of the magnetic field, 
such as the location and evolution of coronal holes, streamer structures, and the HCS; 
however, such an approximation does not predict the density and temperature very accurately \citep{RileyApJ2006}.
In particular, \cite{Vasquez2008ApJ} compared a static tomographic reconstruction of the density
with two polytropic MHD models (Stanford: \cite{Hayes2001ApJ}; and Michigan: \cite{Cohen2007ApJ}
during Carrington rotations 2029.
They found that these polytropic MHD models could reproduced the density values only below 3.5~\Rsun\ and at low latitudes, 
while both models had problems reproducing the correct density in the polar regions. 
%
%
A more recent thermodynamic MHD model uses an improved equation for energy transport in the corona that includes parallel thermal conduction along the magnetic field lines, radiative losses, and parameterized coronal heating. 
This thermodynamic MHD model produces more accurate estimates of plasma density and temperature in the corona \citep{Lionello2009, Riley2011SoPh}.
The electron density estimated by the polytropic MHD model (pMHD/$N_{e}$) for the Carrington rotations 2077 and 2097 are shown in panels~(b) of Figures~\ref{fig_tomo_predsci_2077} and \ref{fig_tomo_predsci_2097}, respectively. Panels~(d) show the radial field calculated by the polytropic MHD model (pMHD/$B_{r}$) for the same Carrington rotations.
\review{The density predicted by the thermodynamic MHD model (tMHD/$N_{e}$) is shown in panels~(c) of Figures~\ref{fig_tomo_predsci_2077} for the Carrington rotation 2077.}
In the left panel, we show the longitude--latitude Carrington map at 3.5~\Rsun;
in the right panel, we show the latitude--radial map obtained by averaging over the longitudes.
The latitudinal locations of the density maximum in pMHD/$N_{e}$ are shown as a green dashed line. The latitudes of the density maximum in tMHD/$N_{e}$ for the thermodynamic MHD model are nearly identical since both models reproduced the general observed configuration of the magnetic field.

It is important to note that the PFSS and the global MHD models require a series of magnetograms providing the nearest central meridian data on the photosphere and covering a full Carrington rotation (27.2753 days), while tomography requires observations of the coronal emission covering only half a rotation, since it relies on optically thin measurements.
Moreover, the photospheric measurements beyond $75^{\circ}$ absolute latitude are not reliable owing to the larger viewing angle with the magnetic field. Therefore, errors in polar field strength estimation at the surface can lead to discrepancies in the modeled magnetic field configuration of the corona. This is especially true during the solar minimum, when the polar fields are the strongest.

\section{Analysis and Comparison}
\label{sec_resu}

\begin{figure}
\begin{tabular}{m{0.1cm} m{0.3\textwidth} m{0.3\textwidth} m{0.3\textwidth}}
& \ \ \ \ \ \ \ \ \ \ Tomography 
& \ \ \ \ Polytropic MHD model 
& Thermodynamic MHD model \\
(a)  & 
\includegraphics[width=0.3\textwidth, trim=0 1.6cm 0 0, clip]{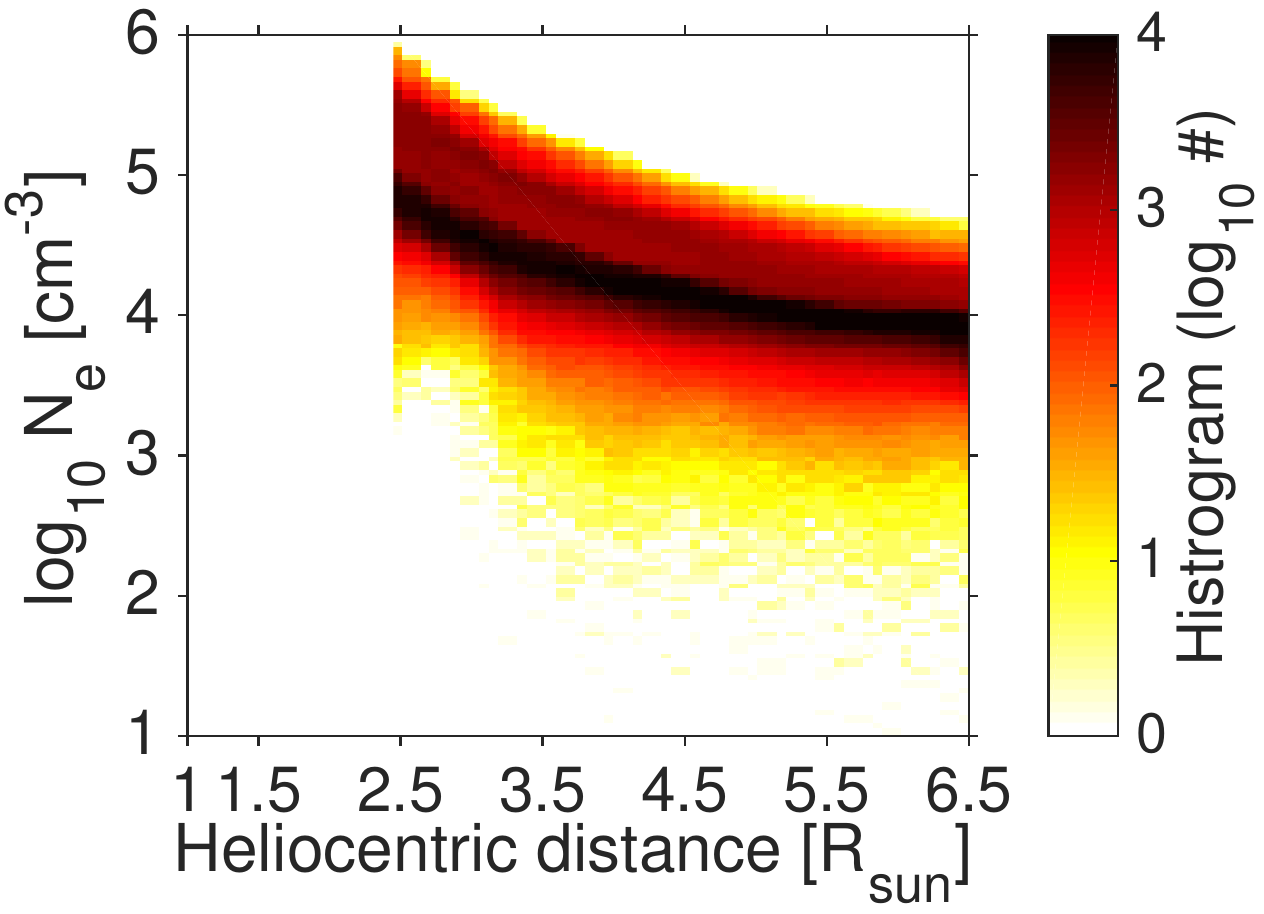} &
\includegraphics[width=0.3\textwidth, trim=0 1.6cm 0 0, clip]{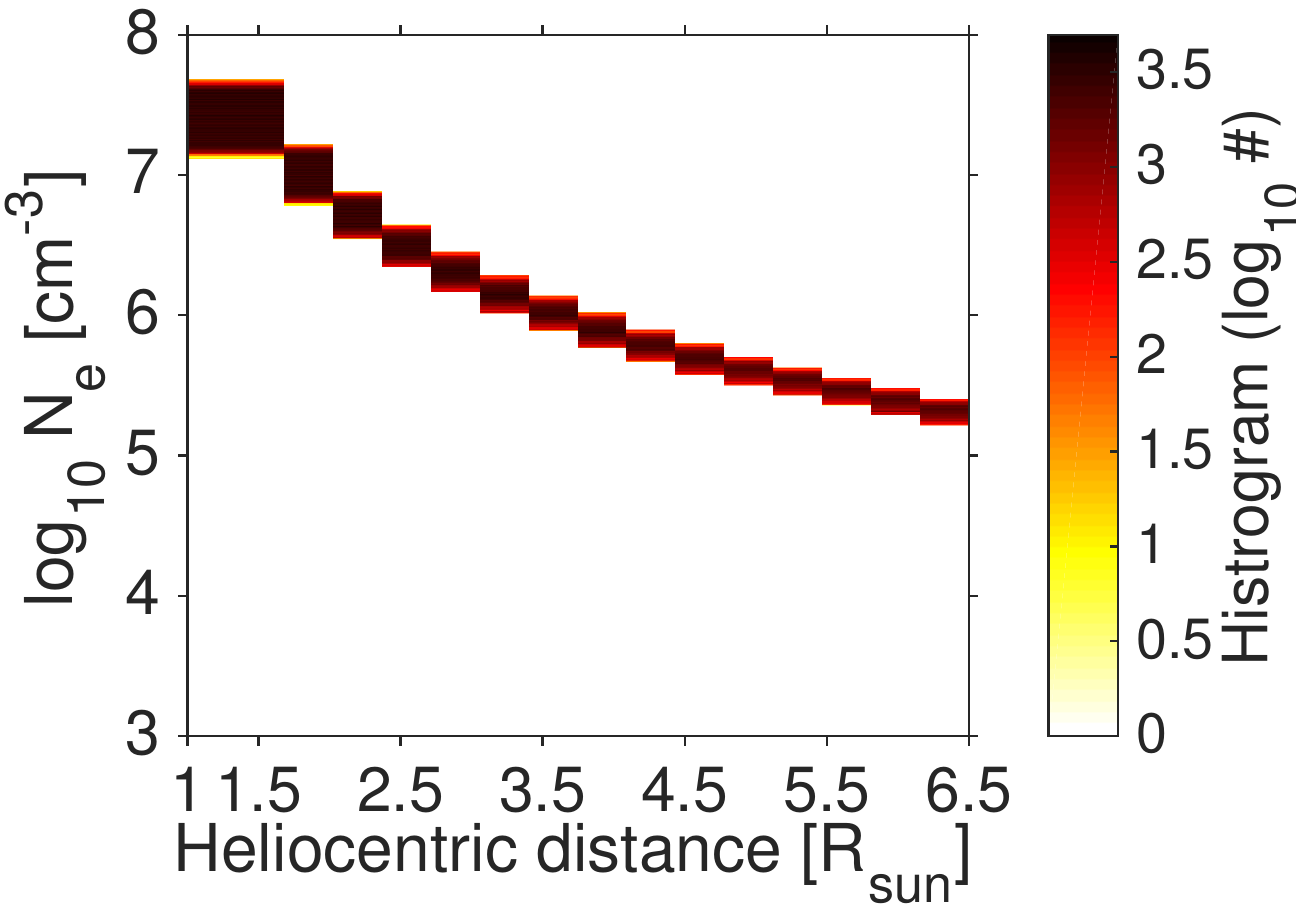} &
\includegraphics[width=0.3\textwidth, trim=0 1.6cm 0 0, clip]{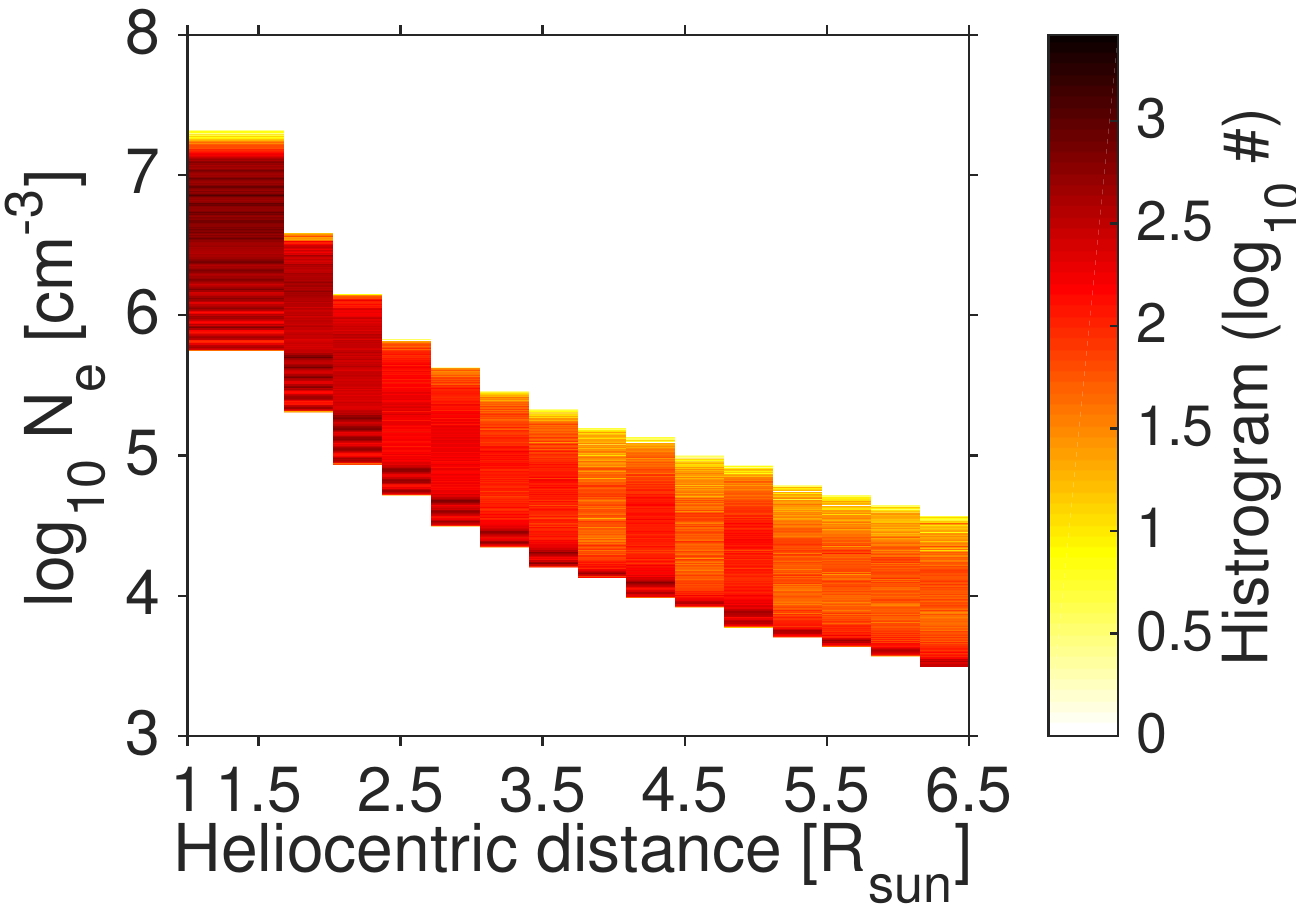} \\
(b)  & 
\includegraphics[width=0.3\textwidth, trim=0 0 0 0, clip]{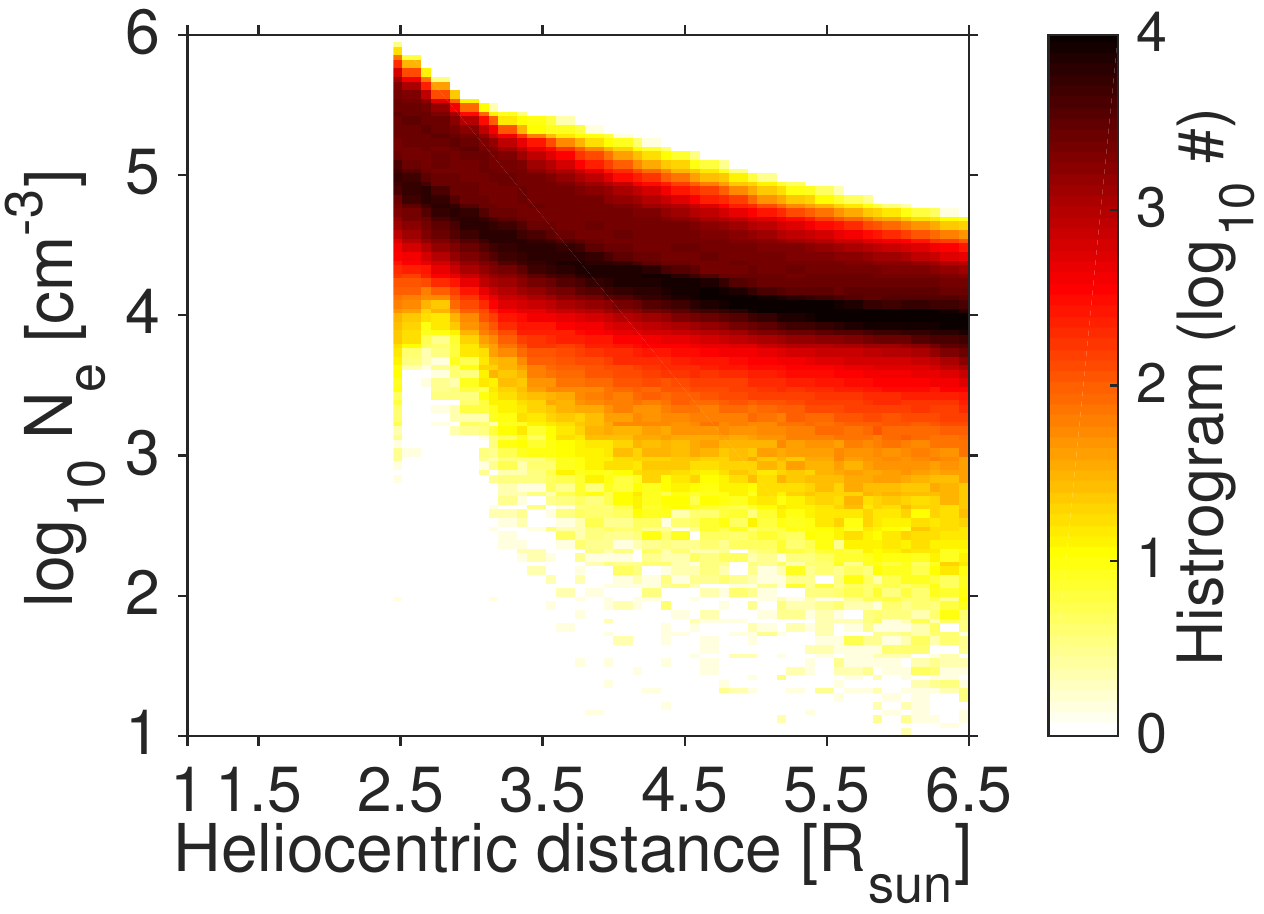} &
\includegraphics[width=0.3\textwidth, trim=0 0 0 0, clip]{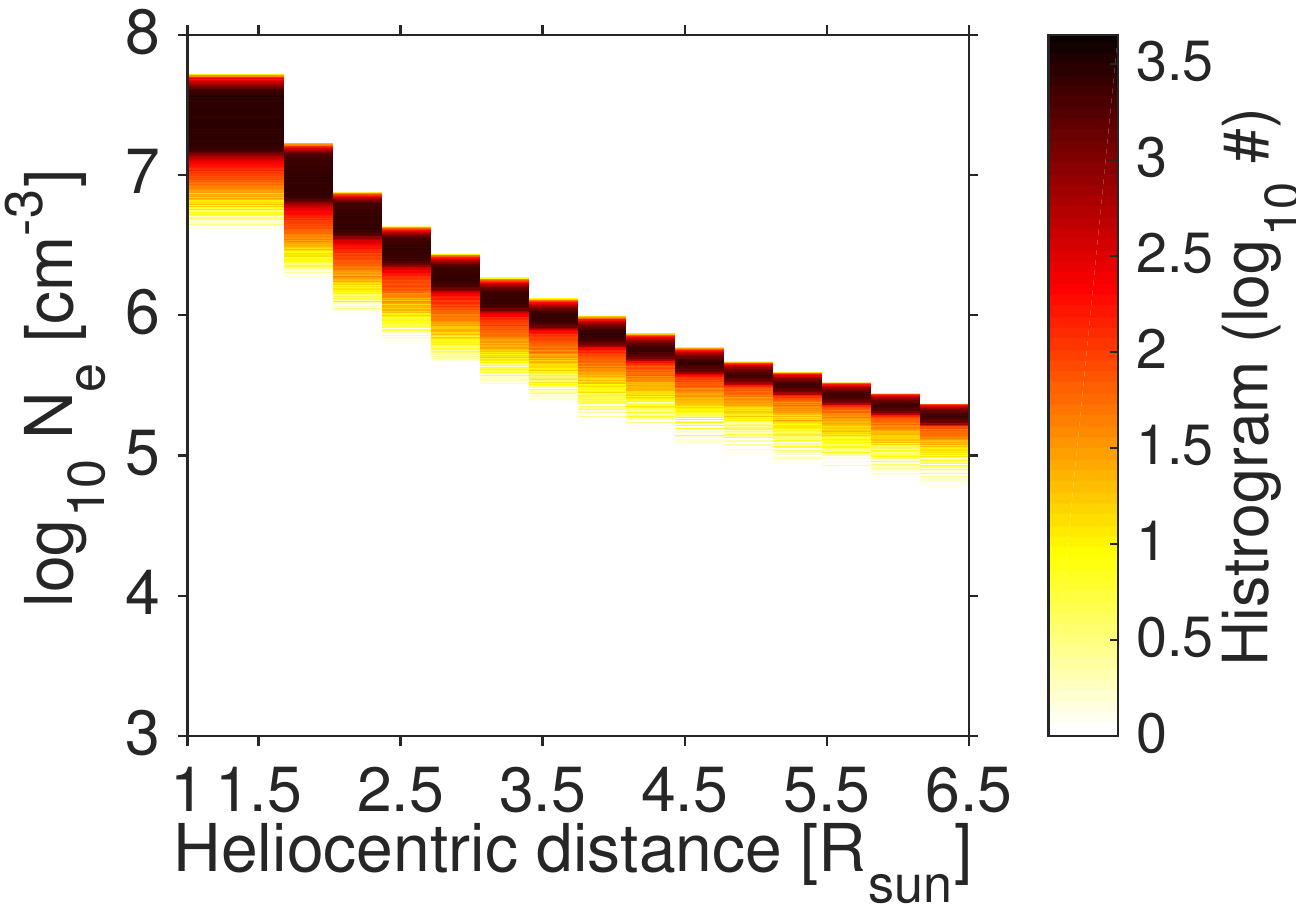} &
\includegraphics[width=0.3\textwidth, trim=0 0 0 0, clip]{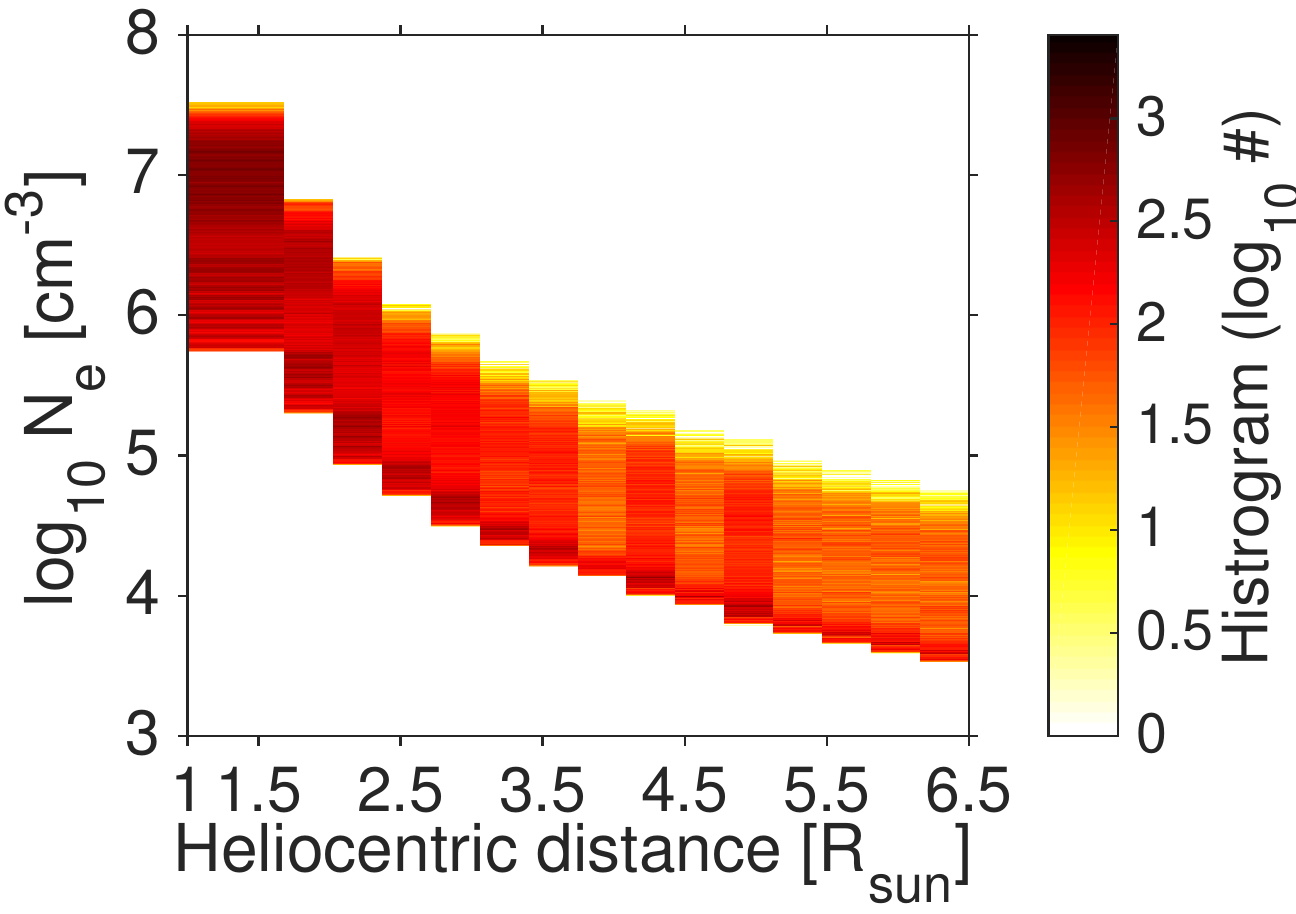}
\end{tabular}
\caption{\review{Histograms of the density distribution for the radial distances.
(a) Tomographic result for 2008 November 21 to December 4, and MHD solutions for Carrington rotation 2077.
(b) Tomography: 2010 June 24 to 2010 July 8, and MHD solutions for Carrington rotation 2098.}
\label{fig_tomo_predsci_histo}}
\end{figure}

%
%
The overall density structure from the MHD models reproduces the essential features of tomography.
Nevertheless, we can see that the results obtained from tomography are more structured, in particular at the poles. 
%
%
The location of the density maximum in pMHD/$N_{e}$ \review{and tMHD/$N_{e}$} (green dashed line, Figures~\ref{fig_tomo_predsci_2077} and \ref{fig_tomo_predsci_2097}) follows nearly exactly the HCS predicted by pMHD/$B_{r}$, which is expected since \review{the models MHD/$B_{r}$ and $N_{e}$ are not independent}.
We observe a clear mismatch between the locations of highest densities from tomography (white dots), 
the PFSS/HCS (black line), 
and the density maximum from the MHD solution (green dashed line).
Previous works showed a limitation of the PFSS model in adequately reproducing some of the observed magnetic structures, in particular when large parts of the solar atmosphere are filled with nonpotential magnetic fields owing to the presence of active regions \citep{Wang2007ApJ_pseudostreamers, Kramar2014SoPh}. Here we show that this is also the case for the HCS predicted by the MHD solutions.
%
%
The density values found for pMHD/$N_{e}$ spread over a narrower range 
(6.3~10$^5$ -- 1.3~10$^6$ cm$^{-3}$) and overestimate the tomography values
(3.1~10$^3$ -- 3.2~10$^5$ cm$^{-3}$) by an order of 
4 for the maximum values and up to 10$^2$ for the minimum values.
Our comparison illustrates the extent to which the plasma parameters predicted by the polytropic MHD model are \review{less realistic compared to the thermodynamic values tMHD/$N_{e}$ (1.9~10$^4$ -- 1.9~10$^5$ cm$^{-3}$)}.
\review{Typical histograms of the density distributions over the radial distances in Figure~\ref{fig_tomo_predsci_histo} show that tomography provides a larger range of density values at every solar radius.}

\subsection{Temporal evolution and radial profiles of the density}

\begin{figure}
\includegraphics[width=0.51\textwidth, trim=1cm 10.8cm 0cm 0cm, clip]{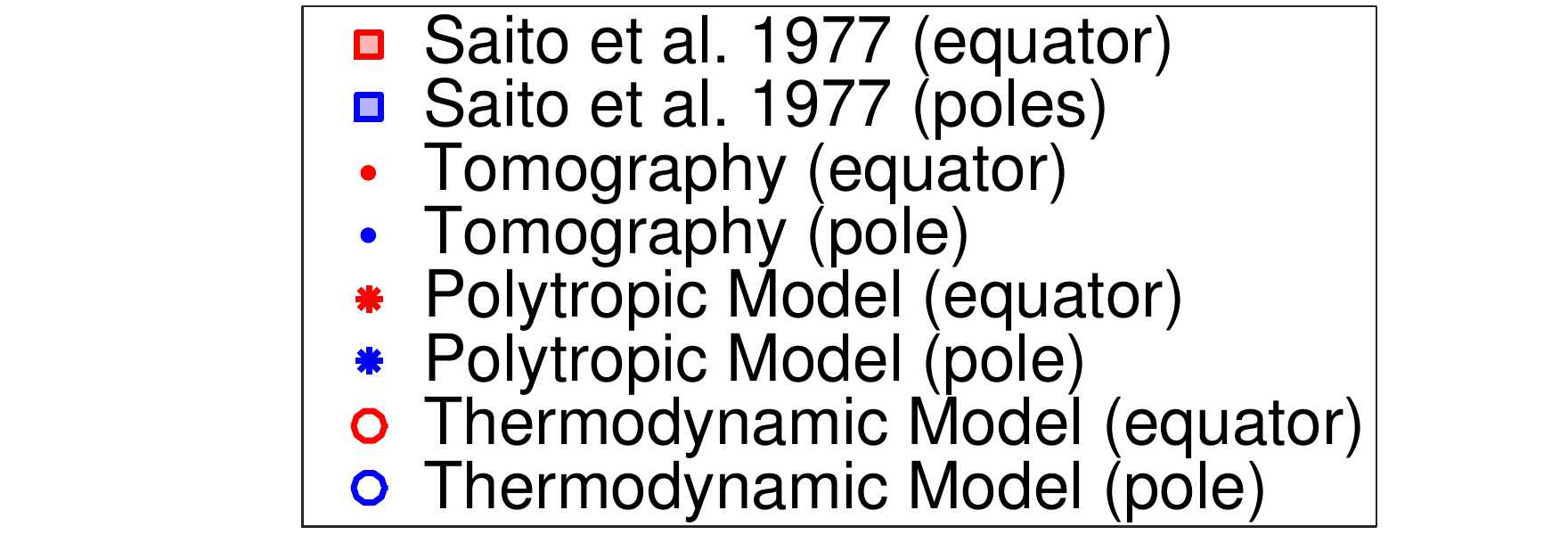}\\
\includegraphics[width=0.8\textwidth, trim=-0.3cm 0.61cm 0cm 0cm, clip]{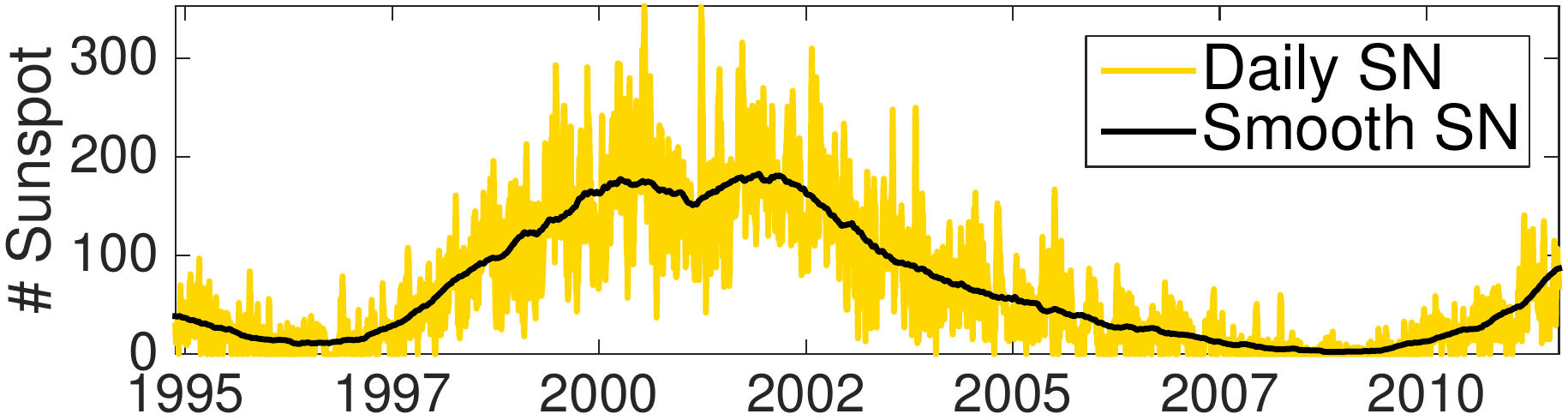}\\
\includegraphics[width=0.8\textwidth, trim=0cm 0cm 0cm 6.2cm, clip]{density_temporal_profile.pdf}
\caption{Top: Sunspot number (SN) \citep[Source:][]{sidc}. 
Bottom: electron density evaluated at 3.5~\Rsun: maximum values of the density are shown in red, and average values at the poles are shown in blue. \label{fig_profile_temp}}
\end{figure}

To investigate the temporal evolution of the density during the two solar cycle minima, we first average over longitude all solutions obtained from tomography, pMHD/$N_e$ and the thermodynamic MHD solutions (tMHD/$N_e$), as it was done for Figures~\ref{fig_tomo_predsci_2077} and \ref{fig_tomo_predsci_2097} right panels of (a) and (b).
We evaluate the ``maximum equatorial'' electron density, $P^{\rm eq}_{N_e}(r,t)$, by taking the maximum density value over the latitudes at each radial distance. We evaluate the \lq{polar\rq} electron density, $P^{\rm pl}_{N_e}(r,t)$, by averaging the density values obtained above 65$^\circ$ and below -65$^\circ$ latitude at each radial distance. %
Figure~\ref{fig_profile_temp} shows the temporal evolution of these densities in the equatorial (red) and polar (blue) regions at a radial distance $r=3.5$~\Rsun.
Since the thermodynamic MHD model is more complex and takes more time to compute, we have fewer data solutions.  

We note first that the temporal evolution of the density distribution from tomography shows a good agreement with the solar cycle; for reference we show the daily sunspot number (SN) and the yearly smooth SN in the top panel of Figure~\ref{fig_profile_temp}. 
In particular, the density values at the equator are found to be lower during the 2008-2010 solar sunspot minimum 
($N_{e}\sim$ 0.8~10$^{5}$ -- 1.1~10$^{5}$ cm$^{-3}$) compared to the 1996-1998 minimum ($N_{e}\sim$ 1.5~10$^{5}$ -- 2.0~10$^{5}$ cm$^{-3}$). 
The minimum in 2008--2009 had 818 days where no sunspot was recorded, and had a yearly smooth SN~$\ge 2.1$, while the minimum in 1996--1997 had only 309 spotless days, with a yearly smooth SN~$\ge 10.4$.
To assess our methodology, we also show the values found by \cite{Saito1977} at $r=3.5$~\Rsun\ (squares) at the equator 
(1.8~10$^{5}$ cm$^{-3}$) and in the polar regions (0.5~10$^{5}$ cm$^{-3}$). 
Saito's densities were evaluated during a previous minimum (solar cycle number 20/21, with 272 spotless days, and a yearly smooth SN~$\ge 16.9$); nevertheless, \cite{Hayes2001ApJ} and \cite{QuemeraisAnA2002} observe good agreement during the first minimum \review{for polar and equatorial regions}. 
At the equator, we consider the higher-density values, while these authors estimate average values of density.
The second minimum, in 2008--2010, shows a lower SN, which reveals how tomography can reproduce the variation of the density distributions that follow the solar cycle. 
\review{At the poles,} the density from tomography is about 40\% that of Saito's for both minima. 
\review{The density models from \cite{Saito1977}, \cite{Hayes2001ApJ} and \cite{QuemeraisAnA2002} are evaluated using the axi-symmetric assumption, which is less reliable than a tomographic inversion.}
During the separation of the K component in the processing and the calibration of the pB images, an overestimation of the F corona and the stray light cannot be excluded, which results in underestimating the K component and thus the estimated density.
\review{On the other hand, these models might also suffer from the missestimation of the background, resulting in incorrect higher values}.
In the future, a new calibration procedure as proposed by \cite{Morgan2015calib} could be used to refine these results. 

As already noted, pMHD/$N_{e}$ overestimates the density found in the tomographic reconstruction by an order of magnitude. On the other hand, tMHD/$N_{e}$ provides more accurate values of the density albeit overestimated at the equator 
(tomo/tMHD $\sim 52$\%) 
and underestimated at the poles 
(tMHD/tomo $\ge 70$\%).
These differences could be linked to the way the equatorial and polar values are computed: recall that the equatorial values correspond to maximum values, while the polar values are averages. It would appear more difficult to obtain a true maximum of a local parcel of plasma with the tomography than it is with the MHD simulation. The lack of resolution at the poles could explain the lower densities in the tMHD model.
No significant time evolution can be observed in pMHD/$N_{e}$, while the tMHD/$N_{e}$ values show time variations that follow the variations in tomography estimates during the minima of the two solar cycles. This is more obvious for the equatorial regions and during the second, more extended solar minimum. Therefore, we conclude that the main variations found in the tomography results are realistic and can be physically interpreted by changes in sunspot activity.

\begin{figure}
\includegraphics[width=0.6\textwidth, trim=-0.8cm 19.2cm 0cm 0cm, clip]{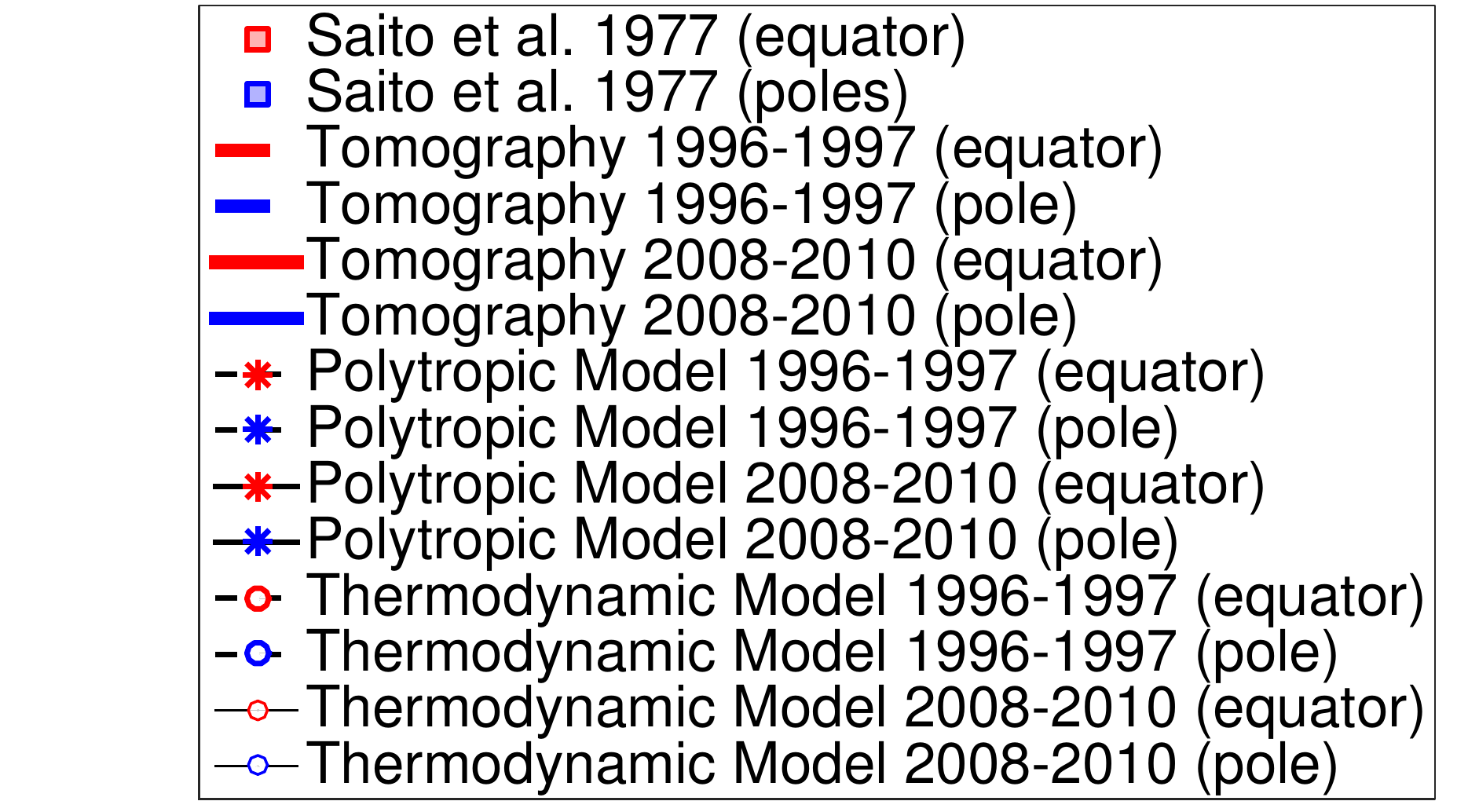}\\
\includegraphics[width=0.80\textwidth, trim=0cm 0cm 0cm 10.5cm, clip]{density_radial_profile.pdf}
\caption{Radial profile of the electron density. Maximum values of the density are shown in red, and average values at the poles are shown in blue. \label{fig_profile_rad}}
\end{figure}

We next study the differences between the two solar minima and estimate radial profiles for the tomographic, pMHD/$N_e$ and tMHD/$N_e$ results.
The {\lq equatorial\rq} radial profiles are obtained by averaging the electron density profiles as follows:
\[\langle P^{\rm eq}_{N_e}(r,t) \rangle_{\rm 1996<t<1997}
\rm{\ \ \ and \ \ \ }
\langle P^{\rm eq}_{N_e}(r,t) \rangle_{\rm 2008<t<2010}.\]
Similarly, we estimate the {\lq polar\rq} radial profiles of the density:
\[\langle P^{\rm pl}_{N_e}(r,t) \rangle_{\rm 1996<t<1997}
\rm{\ \ \ and \ \ \ }
\langle P^{\rm pl}_{N_e}(r,t) \rangle_{\rm 2008<t<2010}.\]
Figure~\ref{fig_profile_rad} shows those radial profiles of the density for the first minimum ($1996<t<1997$) as a dashed line, for the second minimum ($2008<t<2010$) as a solid line, at the equator (red) and at the poles (blue).
Error bars represent the variance of the density values in the tomographic reconstruction over the given time period.
As a reference, we also show the radial profiles found by \cite{Saito1977}.
The general radial profile trends are in reasonably good agreement. Tomography results show profiles slightly more complex, and important changes between the two solar minima are observed. First, at the equator the densities differ by 62\% along the radial profile, showing that the variations between cycles at 3.5~\Rsun\ are found at all radial distances. Second, at the poles the profiles cross at 3.5~\Rsun, showing opposite variations between cycles, below and above this key radial distance, with larger densities in the outer corona during the second deeper minimum.
While the tMHD/$N_e$ profiles at the equator differ by a larger factor of 92\% between the two minima, there is no significant change at the poles. The tMHD/$N_e$ profiles are more consistent with tomography up to 3.4~\Rsun\ and produce lower values at larger radial distances.

\subsection{Location of the highest-density structures}

During the 2008--2010 minimum, comparing the two latitude--radial maps in the declining phase of cycle 23 and the rising phase of cycle 24 (right panels (a) of Figures~\ref{fig_tomo_predsci_2077} and \ref{fig_tomo_predsci_2097}) helps to
show that the denser region, presumably above active regions, spread more in latitude when solar activity increases. It is not obvious that the denser regions always correspond to the helmet streamer.

\begin{figure}
\begin{tabular}{c}
120$^\circ$ \\
\includegraphics[width=0.25\textwidth, trim=0 1.55cm 0 0, clip]{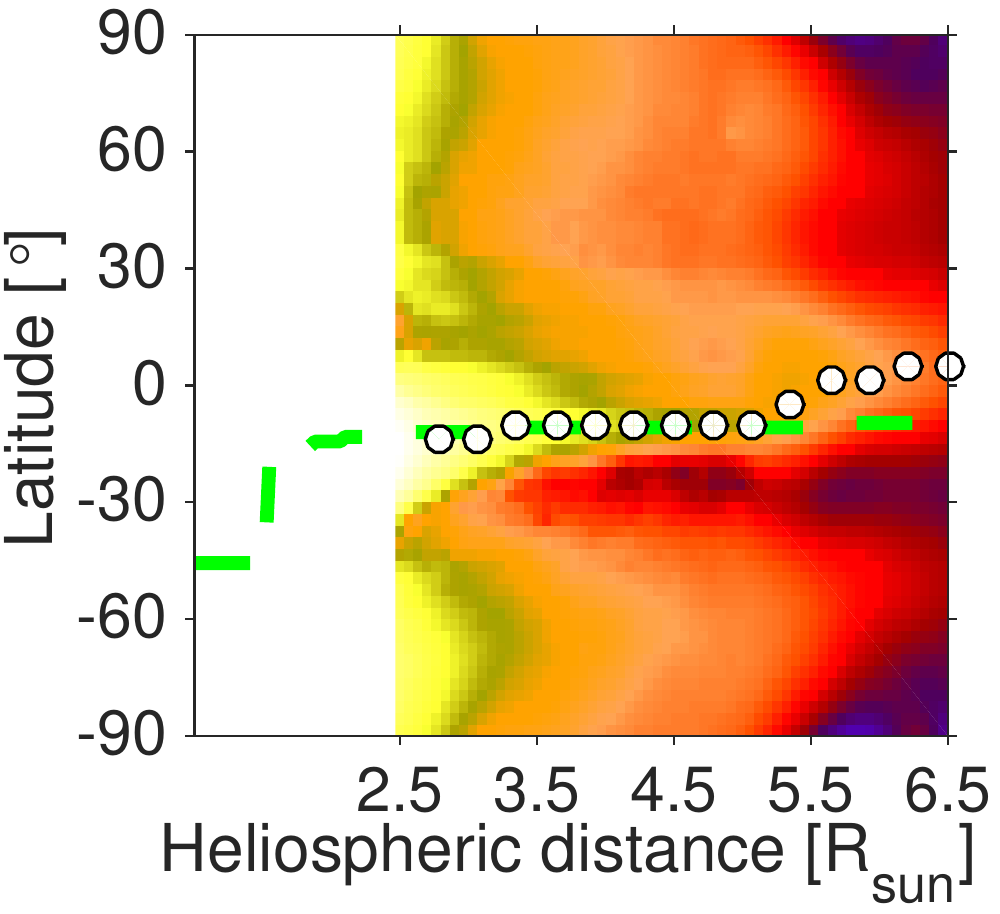} \\
\includegraphics[width=0.25\textwidth, trim=0 1.55cm 0 0, clip]{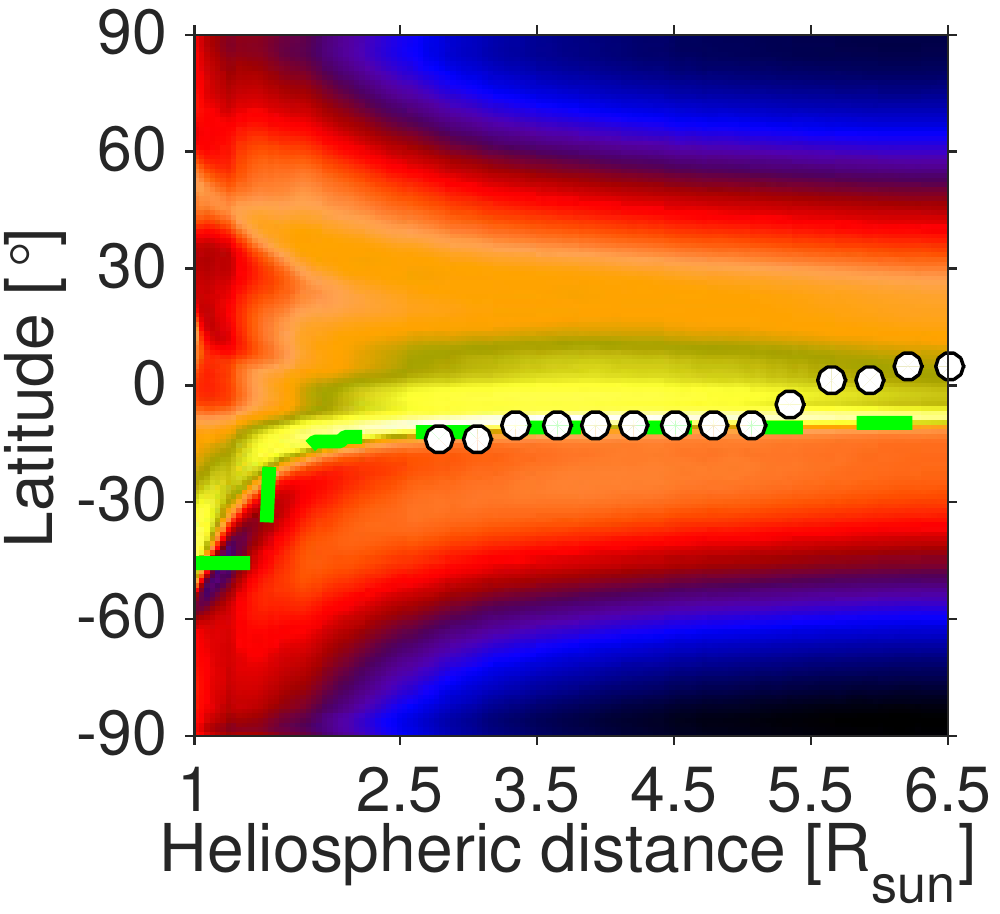}\\
\includegraphics[width=0.25\textwidth, trim=0 0 0 0, clip]{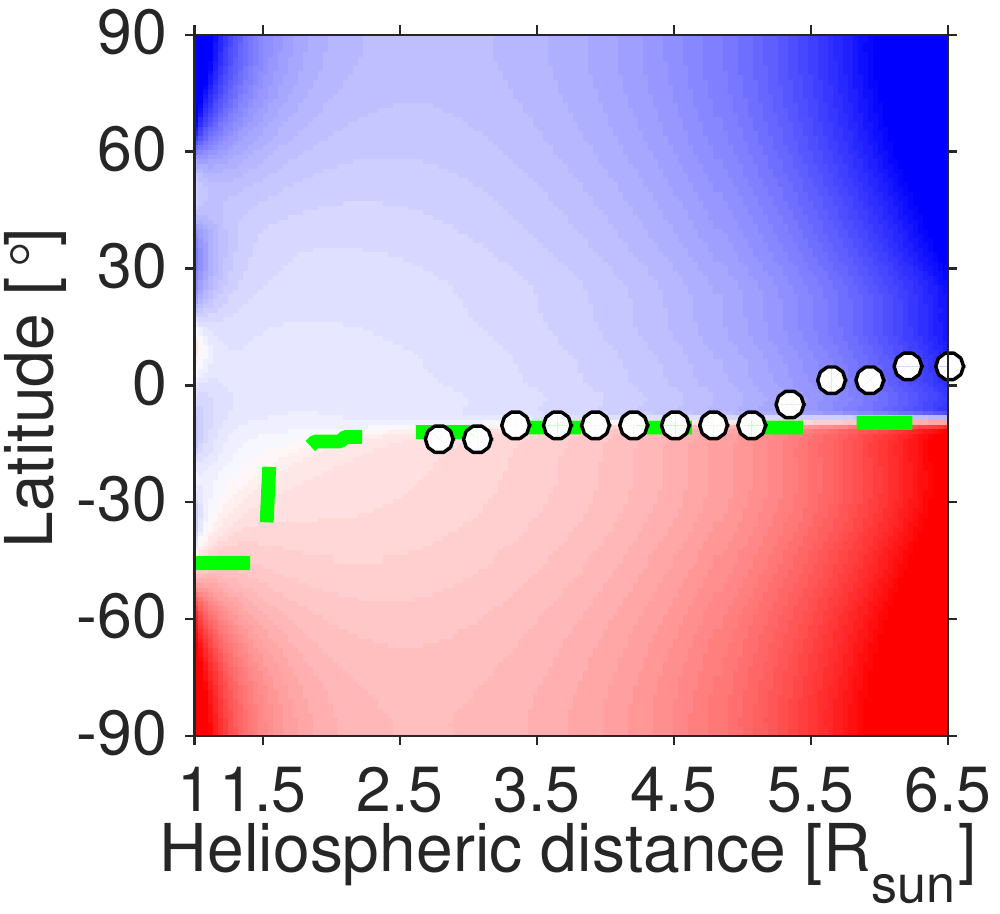}\\
\end{tabular}
\caption{Latitude--radial maps at 120$^\circ$ longitude.
Top to bottom: 
tomographic result (2008 November 21 -- December 4), 
pMHD/$N_e$, and pMHD/$B_r$ (Carrington rotation 2077).
\label{fig_tomo_predsci_2077_LON}}
\end{figure}

\begin{figure}
\begin{tabular}{cc}
90$^\circ$ & 170$^\circ$ \\
\includegraphics[width=0.25\textwidth, trim=0 1.55cm 0 0, clip]{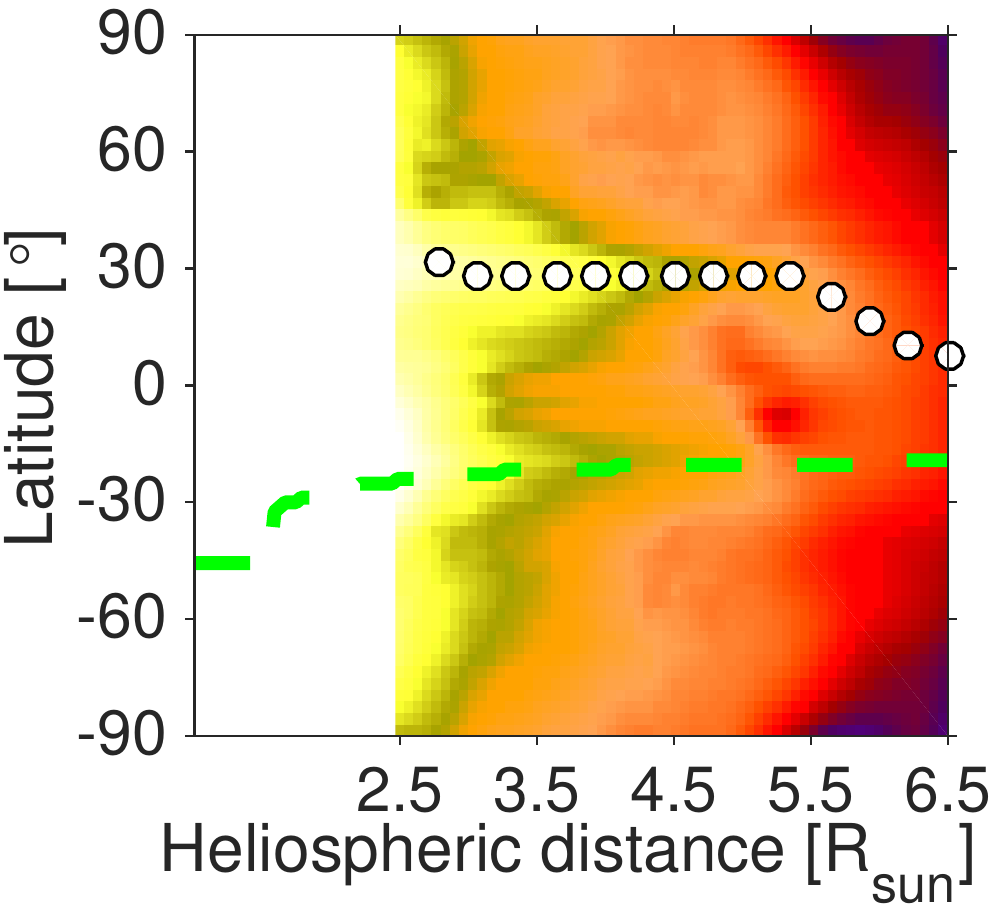} &
\includegraphics[width=0.25\textwidth, trim=0 1.55cm 0 0, clip]{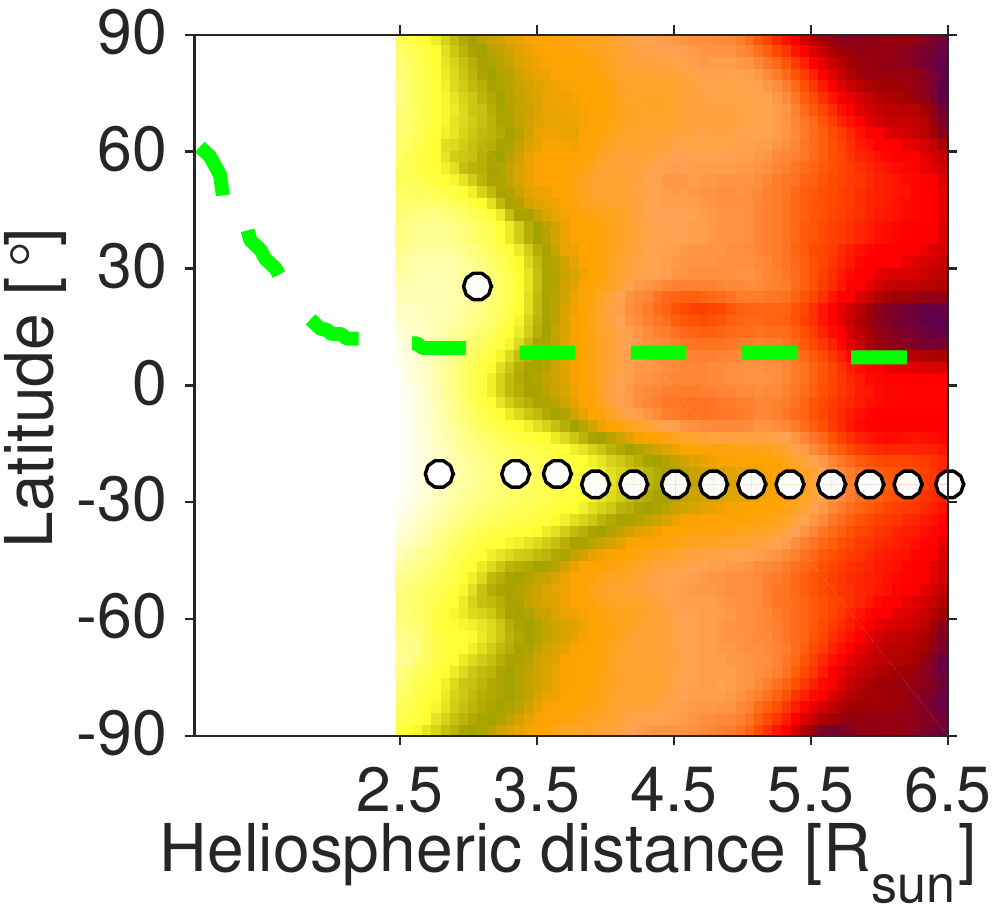} \\
\includegraphics[width=0.25\textwidth, trim=0 1.55cm 0 0, clip]{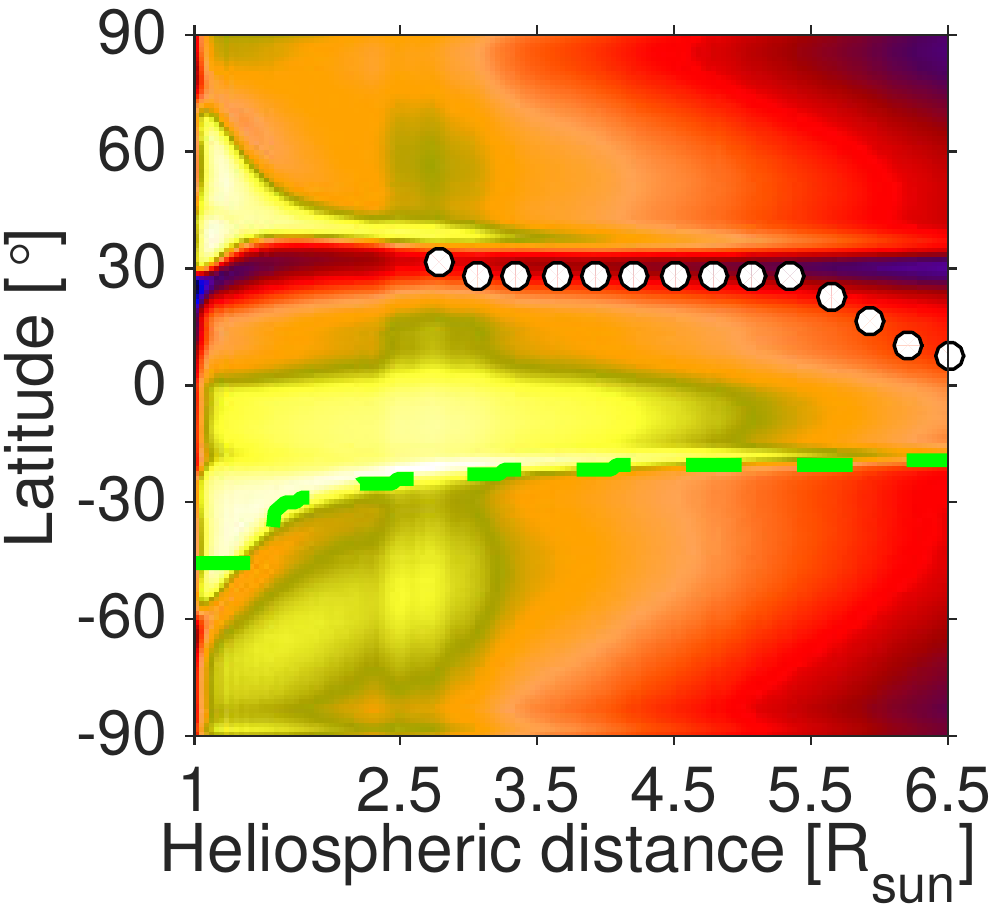} &
\includegraphics[width=0.25\textwidth, trim=0 1.55cm 0 0, clip]{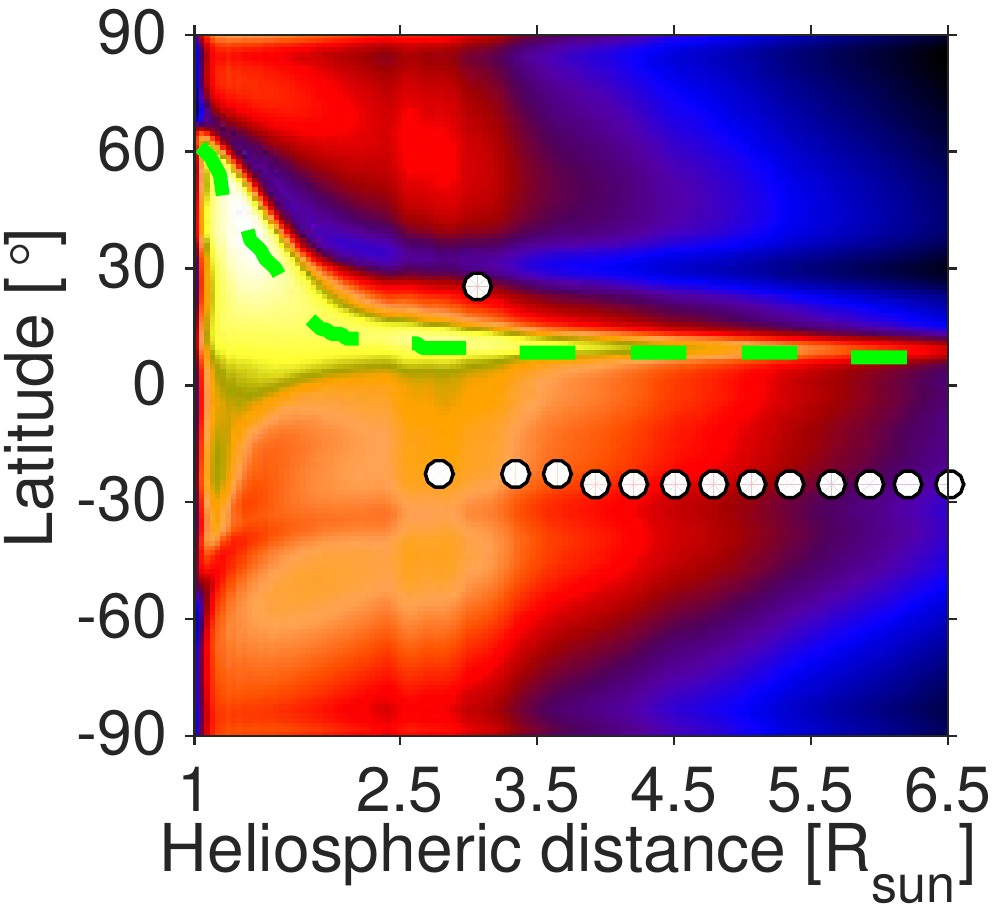} \\
\includegraphics[width=0.25\textwidth, trim=0 0 0 0, clip]{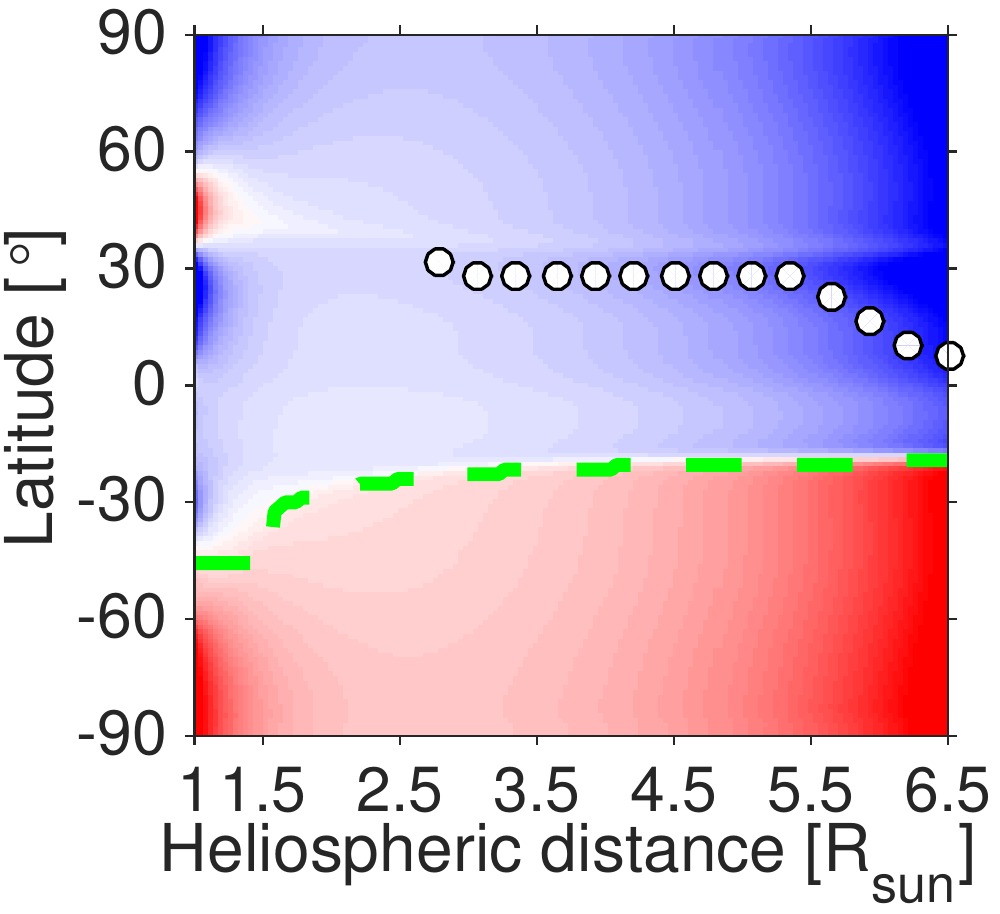} &
\includegraphics[width=0.25\textwidth, trim=0 0 0 0, clip]{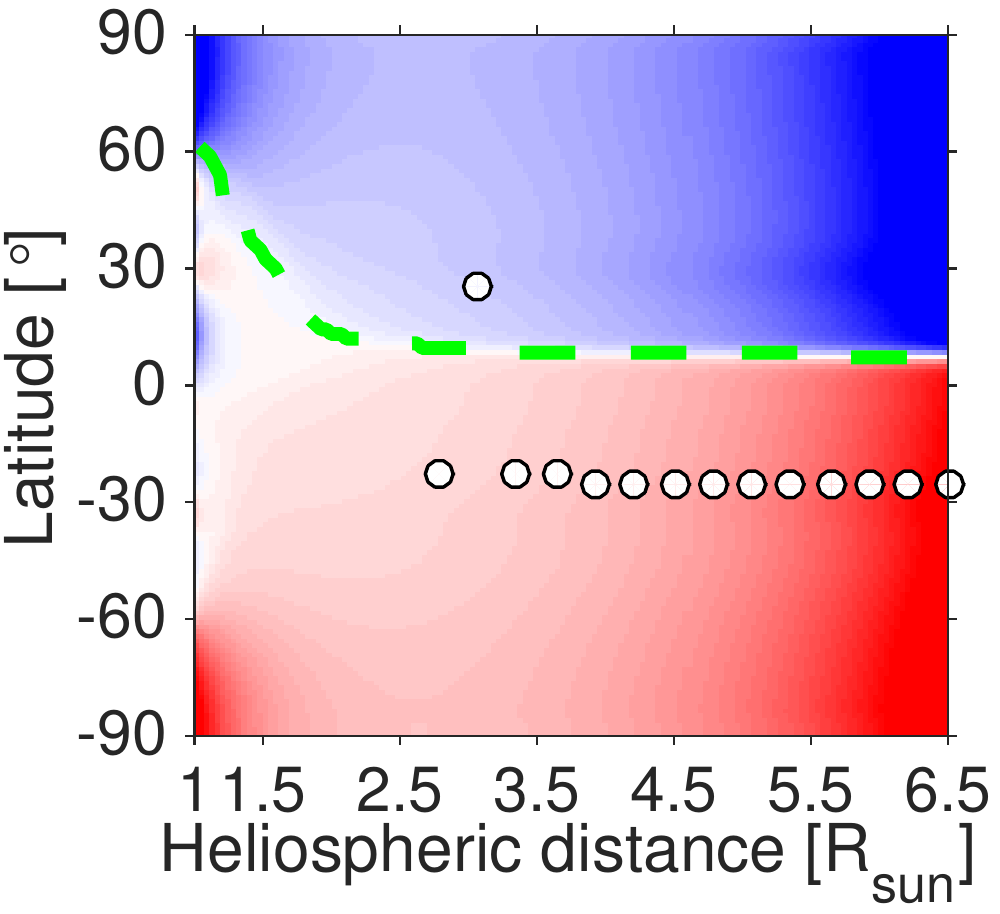}
\end{tabular}
\caption{Latitude--radial maps at 90$^\circ$ and 170$^\circ$ longitude.
Top to bottom: 
tomographic result (2010 Jun 6--20), 
pMHD/$N_e$, and pMHD/$B_r$ (Carrington rotation 2079).
\label{fig_tomo_predsci_2097_LON}}
\end{figure}

We investigate how locations in latitude of the density maximum and the HCS agree or differ with time during the 2008--2010 minimum.
To do so, we estimate the position in latitude of the density maximum in all the tomographic reconstructions and in the pMHD/$N_e$ models for every Carrington rotation from 2065 to 2106.
The latitude of the HCS is extracted both in the PFSS model at the source surface of 2.5~\Rsun\ and in the pMHD/$B_r$ model (as the neutral line where $B_r \simeq 0$) at 1.5 and 3.5~\Rsun.
\review{Panels (a)--(c) of Figure~\ref{fig_lat_temp} show the time evolution of the spread in latitude over all longitudes from the HCS predicted by pMHD/$B_r$ and the higher-density regions in tomography.}
\review{Panels (d)--(g) are longitude--time maps that show the latitudinal locations of the density maximum from tomographic reconstructions, pMHD/$N_e$, and the HCS from pMHD/$B_r$ and PFSS.}
\review{While panels (a) and (b) show that the spread of the HCS predicted by pMHD/$B_r$ is more confined with higher radial distance from 1.5~\Rsun\ to 3.5~\Rsun, the longitude--time maps of pMHD/$B_r$ were found to be the same at 1.5~\Rsun\ and 3.5~\Rsun in panel (f). 
The latitudinal spread of the tomographic highest-density region in panel (c) follows well the predicted HCS spread in panel (b), notably with a widening of the latitude range at the end of 2009.
This change coincides with the rise of the new solar cycle 24 when new sunspots appear at higher latitudes, which results in the streamer belt spanning over higher absolute latitudes.}
As expected, the results from pMHD/$N_e$ and pMHD/$B_r$ in panels (e) and (f) are nearly the same, which show a good agreement between the location of the density maximum and the location of the current sheet predicted by the MHD solution.
We see a reasonably good agreement between the PFSS/HCS in panel (g) and the current sheet predicted by pMHD/$B_r$, which is expected since both are based on the observed LOS measurements of the photospheric magnetic field and uniform characteristic values for the plasma density and temperature as lower boundaries.
The tomographic highest-density structure generally follows the predicted HCS as observed by \cite{Kramar2014SoPh}, especially close to the minimum of solar activity, from 2008 to mid-2009. Here this can be observed thanks to longitudinal drifts with time of the highest-density structures. 
However, this is less clear during the rising phase of the solar cycle 24, towards the end of 2009. 
To investigate this difference, we show latitude--radial planes in the extended minimum and rising phases of cycle 24. Figure~\ref{fig_tomo_predsci_2077_LON} shows latitude--radial planes at longitude 120$^\circ$ of the tomographic reconstruction (2008 November 21 to December 14) and of the pMHD/$N_e$ and  pMHD/$B_r$ solution during Carrington 2077.
In this period of extended minimum, the maximum density in tomography follows the current sheet predicted by pMHD/$N_e$ and pMHD/$B_r$.
On the other hand, Figure~\ref{fig_tomo_predsci_2097_LON} shows two examples of planes taken during the rise of the solar cycle 24 at longitude 90$^\circ$ and 170$^\circ$ during Carrington rotation 2097, where we observe that the maximum density from tomography does not follow the HCS but more likely aligns with a pseudo-streamer.

Therefore, a pseudo-streamer can be found to be denser than a helmet streamer at the same longitude. We conclude that the highest-density structures do not always correspond to the predicted large-scale HCS or its helmet steamer but can follow the locations of pseudo-streamers. Since both structures contribute to the denser regions near the equator, both play a role in the wider spread in latitude as the activity increases.

\subsection{Longitudinal drifts of the highest-density structures}

\begin{figure}
\begin{tabular}{m{0.1cm} m{0.85\textwidth}}
(a)  &
\includegraphics[width=0.505\textwidth, trim=-0.2cm 1.5cm 0 0, clip]{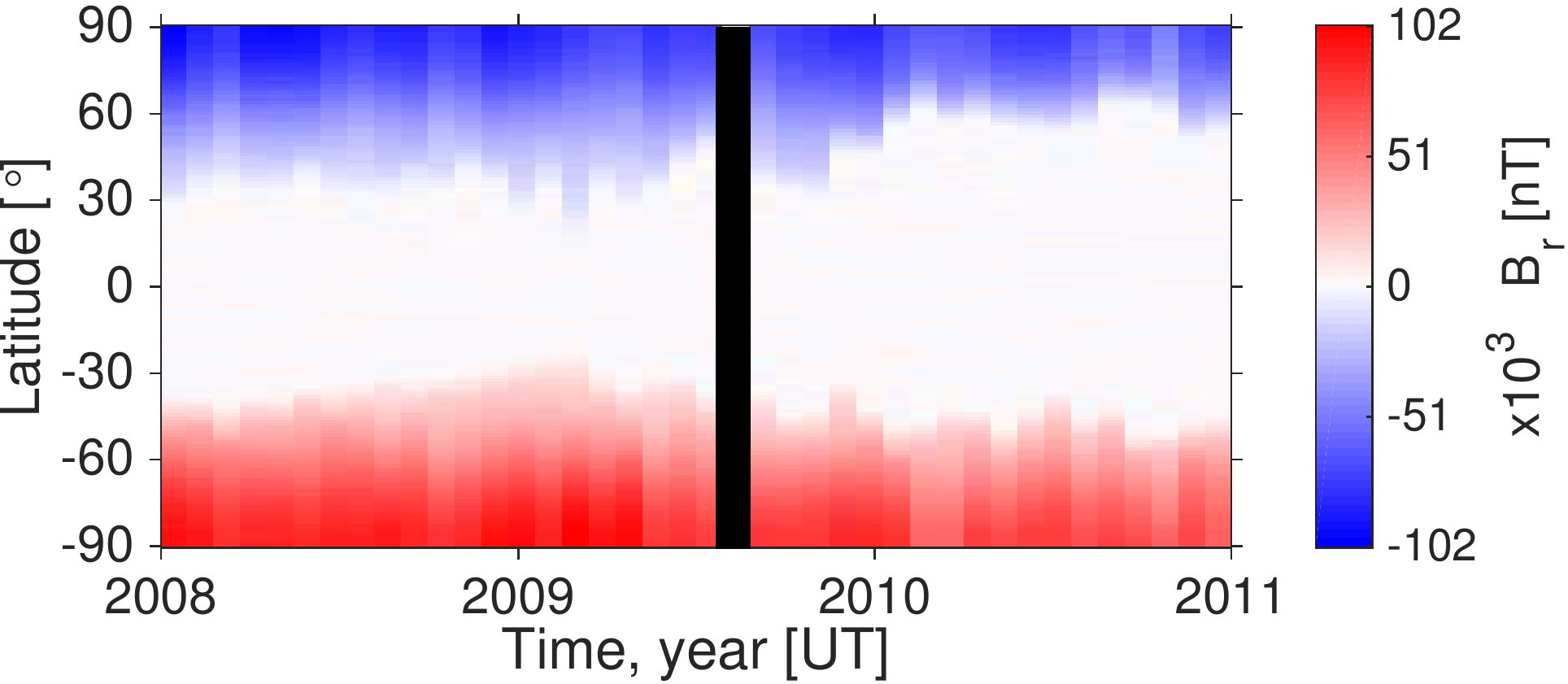}\\
(b)  &
\includegraphics[width=0.505\textwidth, trim=-0.2cm 1.5cm 0 0, clip]{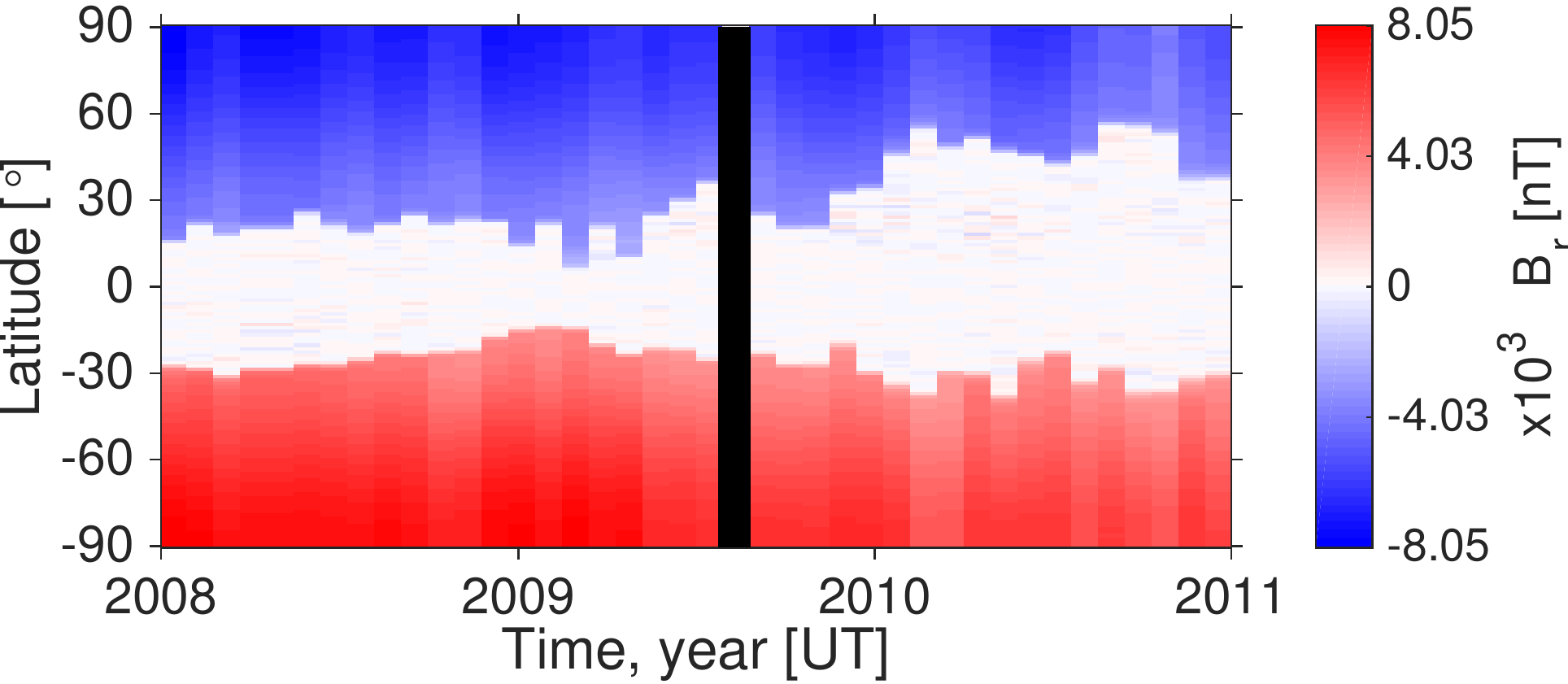}\\
(c)  &
\includegraphics[width=0.505\textwidth, trim=-0.2cm 1.5cm 0 0, clip]{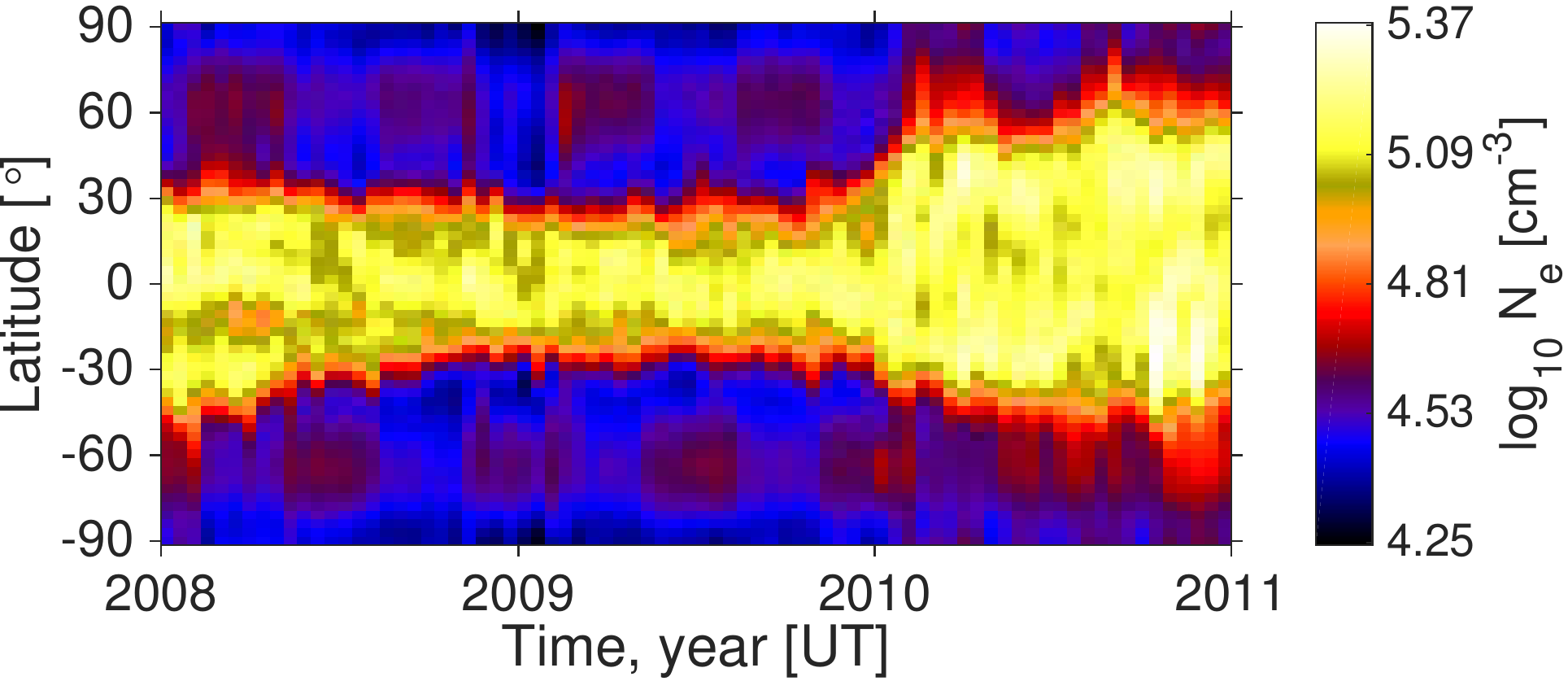}\\
(d)  &
\includegraphics[width=0.49\textwidth, trim=0 1.5cm 0 0, clip]{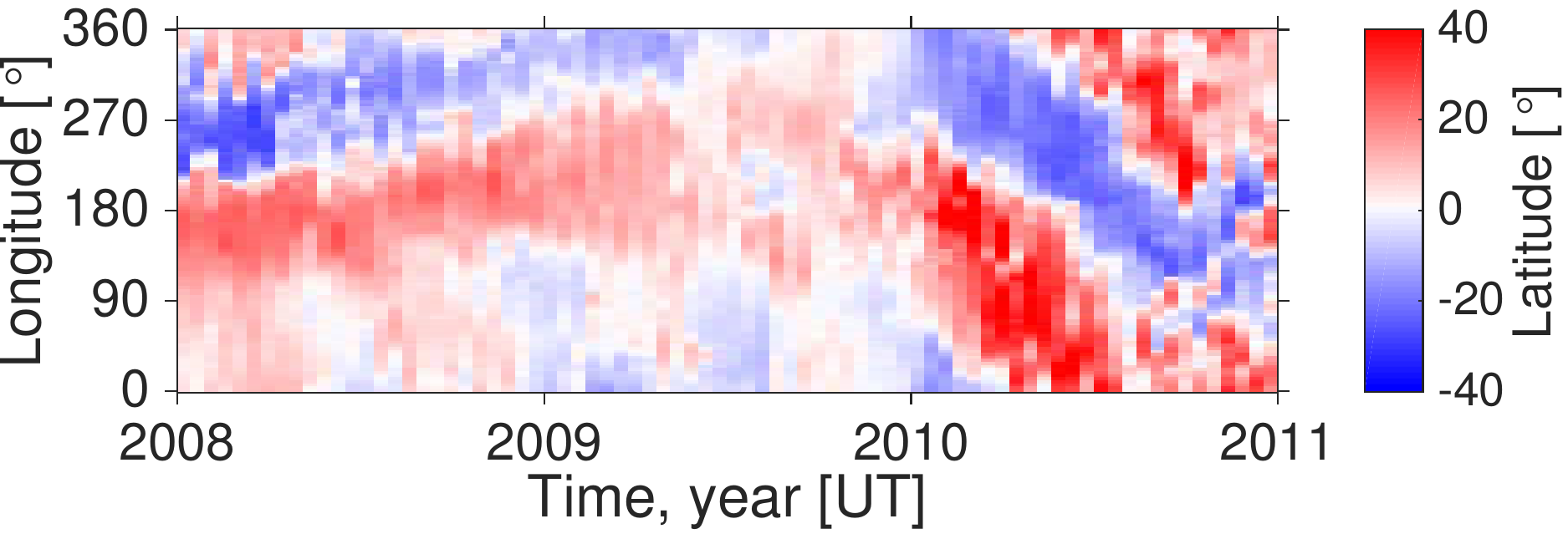}\\
(e)  &
\includegraphics[width=0.49\textwidth, trim=0 1.5cm 0 0, clip]{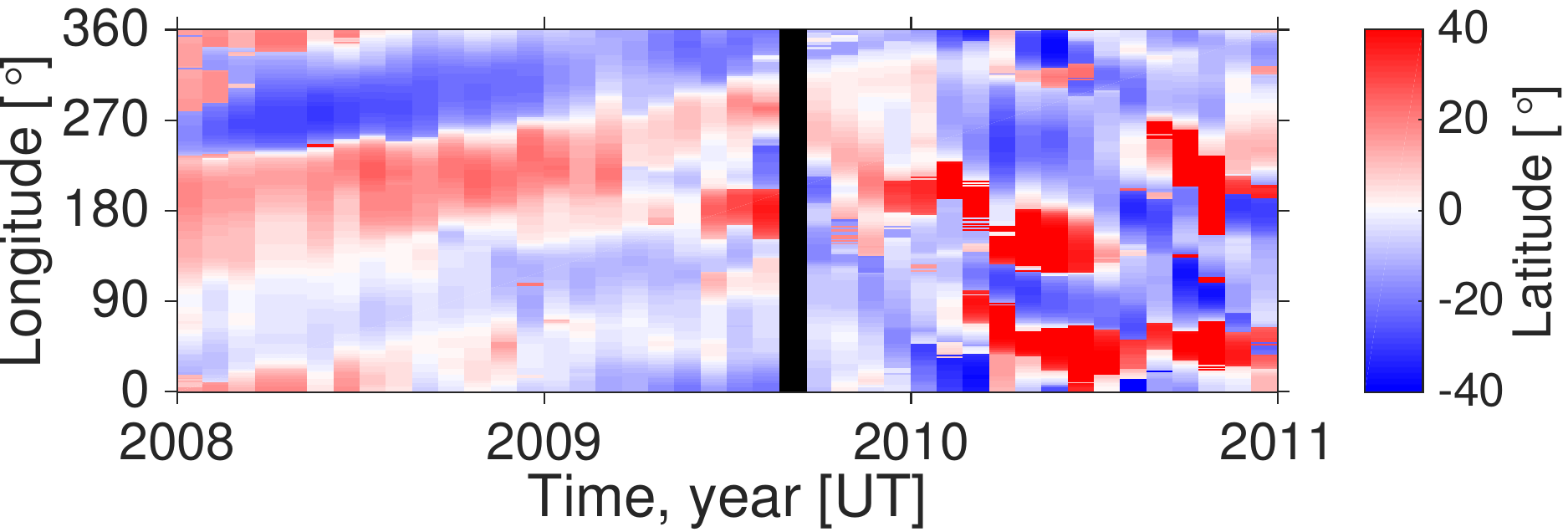}\\
(f)  &
\includegraphics[width=0.49\textwidth, trim=0 1.5cm 0 0, clip]{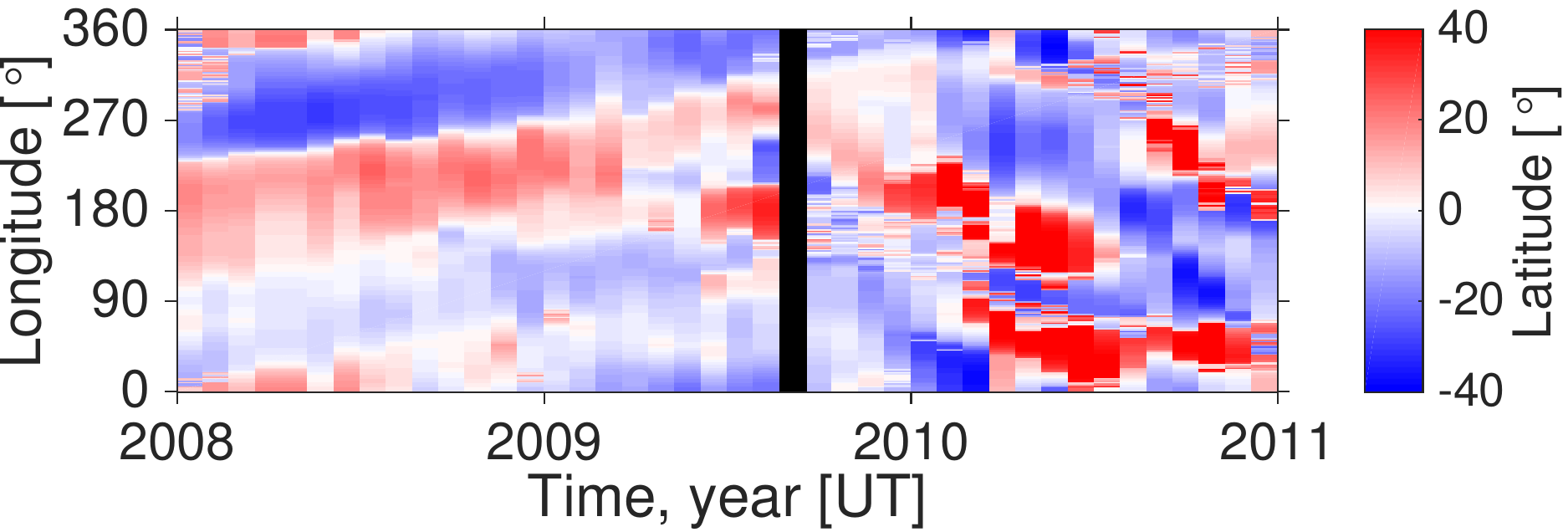}\\
(g)  &
\includegraphics[width=0.49\textwidth, trim=0 0.cm 0 0, clip]{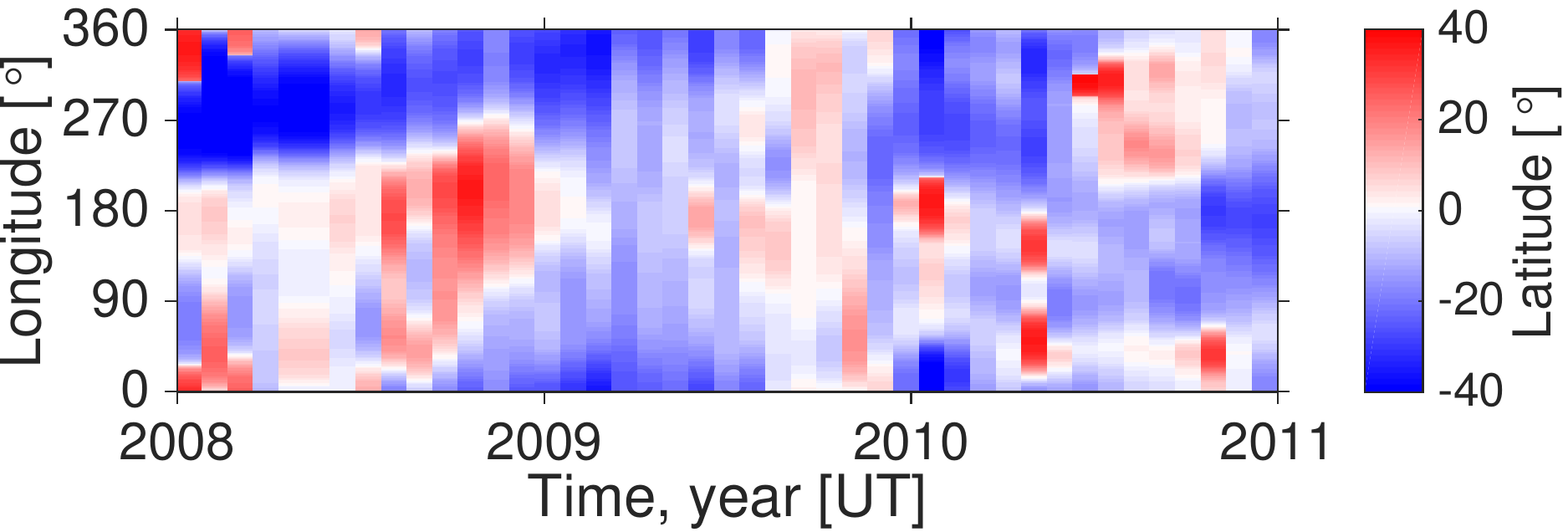}\\
\end{tabular}
\caption{\review{Relations between electron densities and the HCS during the 2008--2010 solar minimum. Latitude--time map of the closeset to zero radial magnetic field predicted by pMHD/Br at (a) 1.5 and (b) 3.5~\Rsun, showing the range of latitudes for the HCS, and (c) latitude--time map of the higher density over longitudes obtained from tomography at 3.5~\Rsun.} Longitude-time of the latitudinal locations of the density maximum at 3.5~\Rsun\ from tomography in (d) and predicted by pMHD/$N_e$ in (e). (f) Latitudinal position of the current sheet predicted by pMHD/$B_r$ at 3.5~\Rsun, and (g) the PFSS/HCS at the source surface (at 2.5~\Rsun). 
The values of the latitude have been truncated above 40$^\circ$ in absolute. The black strips correspond to lacking MHD solutions owing to the missing \textit{SOHO}/MDI synoptic map. \label{fig_lat_temp}}
\end{figure}

Longitudinal drifts with time of \review{coronal structures at 4~\Rsun have been first reported by  \cite{Morgan2011ApJ_LongitudinalDrifts_a, Morgan2011ApJ_LongitudinalDrifts_b}. 
The author measured the rotation rate of structures within specific latitudinal regions (as opposed to the maximum of density studied here) between  -80$^\circ$ and 80$^\circ$ using a back-projection tomographic method. The rotation rates were found to vary considerably between latitudes with values between -3$^\circ$ and 3$^\circ$day$^{-1}$ relative to the Carrington rotation rate.
In Figure~\ref{fig_lat_temp} we observe a longitudinal drift at 3.5~\Rsun of the highest-density structures that} are toward higher longitudes in the extended minimum phase and toward lower longitudes in the rising phase. 
Knowing how the denser regions are spreading in latitudes as the activity increases, we propose that the highest-density structures show a differential rotation well above the surface depending on how they are magnetically connected to the surface.
The tomographic reconstruction method and the MHD models use the approximation of solar Carrington rotation. 
The Carrington rotation rate of 27.2753~days corresponds to the rotation observed near $\pm 30^\circ$ latitudes on the surface of the Sun \citep[e.g.,][]{Snodgrass1990ApJ, Beck2000}. Thus, depending on the latitude of a structure on the surface, its rotation rate, $\omega$ in $^\circ$day$^{-1}$, is larger or smaller than the Carrington rotation rate, 
$\omega_{\rm CR}=13.20^\circ$day$^{-1}$,
\begin{equation}
\omega = \omega_{\rm CR} + \alpha
\label{eq_SolRotation}
\end{equation}
where $\alpha$ is \review{positive} for the structures located between the latitudes $-30^\circ$ and $+30^\circ$ (showing a faster rotation), \review{negative} for the structures above $|\pm 30^\circ|$ (showing a slower rotation), and zero for structures located near $-30^\circ$ or $+30^\circ$.
During the extended minimum, the helmet streamer clustered near the equator. The structure rotated faster than $\omega_{\rm CR}$, and shifted toward the larger Carrington longitudes, resulting in a positive longitudinal drift. From 2008 up to mid-2009, we find a faster rotation rate with
$\alpha\simeq 0.25^\circ$day$^{-1}$, which means that the structure took only about 26.77~days to make a full rotation. 
On the other hand, during the rising phase of the solar cycle, the denser regions spread over latitudes above $|\pm 30^\circ|$, and were associated with a negative longitudinal drift. 
We find a slower rotation rate than $\omega_{\rm CR}$ with 
$\alpha\simeq -0.75^\circ$day$^{-1}$, corresponding to about 28.89 days for a full rotation.
\review{The reversal in rotation rate coincides with the observed sudden extension in latitudes of the structures associated with the rise of solar activity toward the end of 2009 (panel (c) of Figure~\ref{fig_lat_temp}).}

This result shows that the effect of the differential rotation is still visible at 3.5~\Rsun\ although the structure might not spread above $\pm 30^\circ$ at this radial distance. 
It also suggests that the rotation of high-density structures is determined by where they are magnetically connected to the surface of the Sun. 
%

\section{Conclusion}
\label{sec_ccl}

The 3D electron density distribution in the corona was determined for two solar minima: 1996--1997 (solar cycle number 22/23) and 2008--2010 (solar cycle number 23/24) with both an empirical model from a newly time-dependent tomographic method and theoretical models from \review{both polytropic and thermodynamic} MHD solutions. \review{The density distribution is more structured in tomography than in the MHD solutions, in particular in the polar regions. In both MHD models the predicted density distribution is strongly related to the configuration of the calculated magnetic field, and the highest-density structures always follow the HCS. While in tomography the highest-density structures do not always correspond to the predicted current sheet, but can sometimes align with the locations of pseudo-streamers. }

In tomographic reconstructions, the highest density at the equator and the average density at the poles follow the temporal evolution observed in the sunspot cycle. The maximum values in thermodynamic MHD solutions, tMHD/$N_{e}$, along the HCS show also a solar cycle variation, while there is no temporal evolution in polytropic MHD solutions, pMHD/$N_{e}$. This confirms that tMHD/$N_{e}$ are more realistic values than pMHD/$N_{e}$ \citep{Lionello2009}.
The equatorial values of both tomography and tMHD/$N_{e}$ are found to be lower during 2008--2010 compared to 1996--1998, in agreement with differences in the solar sunspot minimum. The tMHD/$N_{e}$ overestimate the tomographic values found at the equator by 52\%, while at the poles the values are consistent up to 3.4~\Rsun\ and then differ. 
At the poles the density from tomography is about 40\% lower compared to \cite{Saito1977} for both minima.

\review{In 2008--2010 the highest-density structures and the HCS predicted by the MHD models show a longitudinal drift, which confirms that the structures do not perfectly follow the Carrington rotation rate, but have a differential rotation also visible well above the surface.
Toward the end of 2009 a drastic change in the rotation rate is observed corresponding to the raising of the solar cycle with the emergence of sunspots at higher latitudes and the spreading of the current sheet across the latitudes. The results suggest that the rotation rate of streamers and pseudo-streamers depends on how the structures are magnetically connected to the surface.}

\review{The following are possibilities for future investigation: 
(1) One could identify the specific rotation rates of latitudinal regions or single structures in the corona independently, as done in the study by \cite{Morgan2011ApJ_LongitudinalDrifts_a}, and contrast the results with an extrapolated radial filed model. 
(2) One could improve} the tomographic method by including the model of the rotation  in the reconstruction, as already done by \cite{dePatoul2013SoPh}, who included the solar differential rotation modeled only at the surface. 
\review{(3) Accurate knowledge of the rotation rate of \review{streamers and pseudo-streamers}} from the surface to higher altitude in the corona could help to better connect the sources of the solar wind to their in situ counterparts \citep[e.g.][]{Foullon2011ApJ,RileyLuhmann2012SoPh}\review{, which can in turn} provide valuable insight for future investigations with \textit{Solar Orbiter} \citep{Muller2013_SolarOrbiter} and \textit{Solar Probe Plus} \citep{Vourlidas2015_ProbePlus}. In particular, \textit{Solar Orbiter} will co-rotate with the Sun and provide images of the polar regions from heliographic latitudes above 35$^\circ$. 
\review{(4)} Ultimately, the time-dependent tomography can be extended to EUV and X-ray ranges to reconstruct also the electron temperature \citep[e.g.][]{Frazin2009ApJ, Vasquez2009SoPh}. It can help to constrain the radial density gradients, base densities, and temperatures of global MHD simulations. Such extensions, combine with the MHD coronal modeling efforts, have the potential to increase the reliability for future space weather forecasting.

\acknowledgments
The authors would like to thank the anonymous reviewer for his/her valuable comments and suggestions to improve the quality of the paper.
J.d.P. is the beneficiary of an AXA Research Fund postdoctoral grant.
C.F. acknowledges financial support from the UK Science and Technology Facilities Council (STFC) under her Advanced Fellowship ST/I003649.
The SOHO/LASCO data used here are produced by a consortium of the Naval Research Laboratory (USA), Max-Planck-Institut for Solar System Research (Germany), Laboratoire d'Astronomie (France), and the University of Birmingham (UK).
SOHO is a project of international cooperation between ESA and NASA.


\bibliographystyle{apj}
\bibliography{article.bib} 
\clearpage


\clearpage

\end{document}